\documentclass[twocolumn]{aastex631}

\usepackage{amsmath,amsxtra}

\begin{document}

\title{Modeling the Multiwavelength Emission of 3C 279 during the 14 years of {\sl Fermi}-LAT Era}

\correspondingauthor{Krishna Mohana A; Lang Cui; Alok C. Gupta}
\email{krishnamohana.mon@gmail.com, krish@xao.ac.cn (KMA); cuilang@xao.ac.cn (LC); acgupta30@gmail.com (ACG)}

\shorttitle{Multiwavelength emission from 3C 279}
\shortauthors{K. Mohana A et al.}

\author[0000-0002-2665-0680]{Krishna Mohana A}
\affiliation{Xinjiang Astronomical Observatory, Chinese Academy of Sciences, 150 Science 1-Street, Urumqi 830011, China}
\affiliation{Aryabhatta Research Institute of Observational Sciences (ARIES), Manora Peak, Nainital 263001, India}
\affiliation{Center for Astrophysics, Guangzhou University, Guangzhou, 510006, People's Republic of China}

\author[0000-0002-9331-4388]{Alok C. Gupta}
\affiliation{Xinjiang Astronomical Observatory, Chinese Academy of Sciences, 150 Science 1-Street, Urumqi 830011, China}
\affiliation{Aryabhatta Research Institute of Observational Sciences (ARIES), Manora Peak, Nainital 263001, India}

\author[0000-0002-5929-0968]{Junhui Fan}
\affiliation{Center for Astrophysics, Guangzhou University, Guangzhou, 510006, People's Republic of China}
\affiliation{ Greater Bay Brand Center of the National Astronomical Data Center, Guangzhou 510006, People’s Republic of China}
\affiliation{Astronomy Science and Technology Research Laboratory of Department of Education of Guangdong Province, Guangzhou, 510006, China}

\author[0000-0003-2011-2731]{Narek Sahakyan}
\affiliation{ICRANet-Armenia, Marshall Baghramian Avenue 24a, Yerevan 0019, Armenia}
\affiliation{ICRANet, P.zza della Repubblica 10, 65122 Pescara, Italy}
\affiliation{ICRA, Dipartimento di Fisica, Sapienza Universita` di Roma, P.le Aldo Moro 5, 00185 Rome, Italy}

\author[0000-0003-1784-2784]{Claudia M. Raiteri}
\affiliation{INAF, Osservatorio Astrofisico di Torino, via Osservatorio 20, I-10025 Pino Torinese, Italy}

\author[0000-0003-0721-5509]{Lang Cui}
\affiliation{Xinjiang Astronomical Observatory, Chinese Academy of Sciences, 150 Science 1-Street, Urumqi 830011, China}
\affiliation{Key Laboratory of Radio Astronomy and Technology (Chinese Academy of Sciences), A20 Datun Road, Chaoyang District, Beijing, 100101, China}
\affiliation{Xinjiang Key Laboratory of Radio Astrophysics, 150 Science 1-Street, Urumqi 830011, China}

\author[0000-0002-0393-0647]{Anne L\"ahteenm\"aki}
\affiliation{Aalto University Mets\"ahovi Radio Observatory, Mets\"ahovintie 114,02540 Kylm\"al\"a, Finland} 
\affiliation{Aalto University Department of Electronics and Nanoengineering, P.O. BOX 15500, FI-00076 AALTO, Finland}

\author[0000-0003-0685-3621]{Mark Gurwell}
\affiliation{Center for Astrophysics, Harvard $\&$ Smithsonian, Cambridge, MA 02138, USA}

\author[0000-0003-1249-6026]{Merja Tornikoski}
\affiliation{Aalto University Mets\"ahovi Radio Observatory, Mets\"ahovintie 114,02540 Kylm\"al\"a, Finland}

\author[0000-0003-1743-6946]{Massimo Villata}
\affiliation{INAF, Osservatorio Astrofisico di Torino, via Osservatorio 20, I-10025 Pino Torinese, Italy}




\begin{abstract}
We report the results of our long-term multiwavelength spectral energy
distribution (SED) study on the flat spectrum radio quasar 3C 279 during the $\sim14$ years (2008--2022) of {\sl Fermi}-LAT  (Large Area Telescope) observing period. The {\sl Fermi}-LAT data were complemented with data in other wavebands obtained from {\sl Swift}-XRT/UVOT, Whole Earth Blazar Telescope (WEBT), along with other optical and radio data from several observatories. Different activity states were identified from the weekly binned $\gamma$-ray light curve, and it was possible to create 168 high-quality and quasi-simultaneous broadband SEDs. We modeled the SEDs using a one-zone leptonic scenario, including the emission region outside the broad-line region (BLR), involving synchrotron, synchrotron self-Compton, and external Compton mechanisms. Such extensive broadband modeling is essential for constraining the underlying multiwavelength radiative mechanisms in the 3C 279 jet and permits to estimate the physical parameters and explore their evolution in time. Our SED modeling study suggests that the increase in the Doppler beaming factor along with the variation of the emitting electrons is the cause for the flares in this source. The multiwavelength emission of 3C 279 was found to be well explained by the scenario in which the emission region is outside the BLR at a distance of $\sim6.42\times10^{3} R_S$. However, for two of the very bright $\gamma$-ray states, the emission region was found to be close to the outer boundary of the BLR at a distance of $\sim1.28\times10^{3} R_S$ from the central black hole.


\end{abstract}



\section{Introduction} \label{sec:intro}
Blazars are a class of active galactic nuclei (AGNs) that exhibit
multiwavelength (MW) emission dominated by the non-thermal radiation originating from their relativistic jet pointing close to the direction of the Earth, and thus is highly Doppler boosted \citep{urry1995}. 
The radio to high energy (HE; $>100$ MeV) and very high energy $\gamma$-ray emission from blazars is highly variable and the time scales of such variations ranging from all the timescale from minutes to years \citep[e.g.,][and references therein]{boettcher2004,foschini2013,orienti2014,boettchergalaxy2019}. From the observational perspective, blazars are sub-classified as flat spectrum radio quasars (FSRQs) and BL Lacerate objects (BL Lacs).  
FSRQs have emission line equivalent widths 
(EW) \textgreater $5$ \AA, whereas BL Lacs have EW \textless $5$ \AA ~\citep{stocke1991,stickel1991}.
According to the classification proposed by  \citet{ghisellini2011}, 
which is based on the broad-line region (BLR) luminosity, FSRQs 
have greater BLR luminosity ($L_{\mbox{BLR}}/L_{\mbox{Edd}} >5\times10^{-4}$) than BL Lacs, where $L_{\mbox{Edd}}$ is the Eddington luminosity for the supermassive black hole at the center of the AGN. \\

The broadband spectral energy distribution (SED) of blazars 
spans the radio to $\gamma$-ray bands and shows a double hump structure \citep{fossati1998}. 
The matter composition inside the blazar jets and the dominant particle 
population responsible for the observed emission in both humps is still not certain \citep{boettchergalaxy2019}. However, most blazar studies consider the ultra-relativistic
electrons or positrons (leptonic scenario) and, in some cases, 
protons (hadronic scenario) to be the primary emitters 
\citep{bottcher2013ApJ}. 
In either scenario, the first hump of the SED, 
which peaks near the optical/UV band is explained by synchrotron 
emission \citep{blumenthal1970RvMP} originating from leptons in the relativistic jet \citep{abdo2010b,bottcher2013ApJ,boettchergalaxy2019,bhattacharya2021}. Furthermore, depending on the location of the synchrotron component peak ($\nu_{\rm s}$), the blazars are classified as low synchrotron peak objects (LSP), when $\nu_{\rm s}<10^{14}$ Hz, intermediate synchrotron peak objects (ISP) and high synchrotron peak objects (HSP), when $10^{14}<\nu_{\rm s}<10^{15}$ Hz and $\nu_{\rm s}>10^{15}$ Hz, respectively \citep{abdo2010b,fan2016,yang2022}.\\

In addition to the synchrotron radiation produced, it is also possible that these relativistic charged particles in the jet can absorb the synchrotron radiation and get energized. This phenomenon is referred to as “synchrotron self-absorption” (SSA;~\citealt{konigl1981ApJ}).
The second hump peaks at the hard X-ray/$\gamma$-ray band and is because of the inverse Compton (IC;~\citealt{blumenthal1970RvMP}) 
emission process produced due to the up-scattering of low energy photons by the energetic jet electrons.
If the low energy photons are produced by the 
synchrotron emission from the jet electrons, the process 
is called synchrotron self-Compton (SSC;~\citealt{marscher1985}). 
The seed photons for IC processes can also come from outside the jet 
\citep{begelman1987}. $\gamma$-ray emissions from FSRQs are 
dominated by up scattering of low energy photons from the accretion disk 
\citep{dermer1993}, BLR \citep{sikora1994}, a dusty torus \citep[DT;][]{bejowski2000}, etc. 
such a process is called an external Compton (EC) process \citep{dermer1992,ghisellini2009}. Alternatively, in the hadronic models the high energy emission is attributed to the interactions of relativistic jet protons in the presence of magnetic field (protron synchrotron;~\citealt{mucke2001APh}) and/or with soft photon field (photopion cascade;~\citealt{mannheim1992}).\\

The SEDs of FSRQs and BL Lacs exhibit different characteristics.
The HE component in the FSRQs are modified due to the strong external photon fields, whereas in case of BL Lacs these photon fields are weak or absent.
In the comoving frame of
the jet, these external photons are relativistically boosted and surpass over the internal synchrotron photon fields, giving rise to the EC component \citep[e.g.,][]{2017MNRAS.470.2861S,2018ApJ...863..114G}.
Hence, in FSRQs, the second component's luminosity is typically greater, indicating a higher level of Compton dominance \citep{sikora1994}. The distribution of the upscattering photons and the location of the emitting region from the central black hole play crucial role in determining the shape of this second component.
Also, the SED parameters or the features in the $\gamma$-ray spectrum like break or cut-off can
constrain the location of the emitting region \citep[e.g.,][]{Poutanen2010ApJ,aleksi2011}.\\

3C 279 is one of the brightest blazar in the $\gamma$-ray band and has been
classified as a FSRQ at a redshift of $z=0.5362,$ \citep{marziani1996ApJS} with a black hole with mass in the range of (3--8) $\times \ 10^{8}M_\odot$ \citep{Gu2001,woo2002}.
The $\gamma$-ray emission from this source was first detected by 
``Energetic Gamma-Ray Experiment Telescope'' (EGRET) on-board ``{\sl Compton Gamma Ray Observatory''} 
\citep[CGRO;][]{hartman1992ApJ}. 
The source has been continuously monitored by the 
Large Area Telescope (LAT) on board the {\sl Fermi} satellite in the energy range 0.1–300 GeV since its launch in 2008.
During the last decade, utilizing the {\sl Fermi}-LAT and other multiwavelength observations, several intensive studies were reported to understand the 
broadband SEDs of 3C 279 during different activity states. 
\citet{abdo2010c} presented compelling evidence demonstrating a correlation between $\gamma$-ray flares and the angle of optical polarization. This finding strongly supports the standard one zone leptonic model.
The presence of additional $\gamma$-ray emission processes \citep{aleksi2011AA,sahayanathan2012MNRAS} has sparked numerous discussions, thanks to the variability at minute timescale reported by earlier studies \citep{paliya2015ApJa,paliya2015ApJb,paliya2016ApJ,ackermann2016ApJ, shukla2020NatCo,wgg2022PASP}, strong Compton dominance \citep{pittori2018ApJ}, and the discovery of the source in the very high energy (VHE) band \citep{magic2008Sci}.\\

The validity of one-zone leptonic models is also questioned by these findings, thus prompting only a limited number of studies to employ alternative models. 
The study conducted by \citet{ackermann2016ApJ} examines the synchrotron source of $\gamma$-rays originating from a jet that is dominated by magnetic forces. 
A model based on a clumpy jet consisting of compact plasmoid strings has been examined by \citet{vittorini2017ApJ}.
The observed HE emission has been investigated by the High Energy Stereoscopic System (H.E.S.S) collaboration, who employed both hadronic and lepto-hadronic models in their explanations \citep{hess2019A&A}.
\citet{rani2018ApJ} have investigated the correlation between parsec scale jet activity and broadband emission through radio observations. They have put forth the argument that the source may contain multiple locations of energy dissipation. Extensive research has been conducted in the literature to investigate the location of the emission region for $\gamma$-ray in the jet. To avoid significant internal absorption, it is necessary for the emission region to be located outside the BLR, as evidenced by the detection of VHE emission from 3C 279.
Recent research on numerous instances of flaring events from the source has indicated that the emission region is situated beyond the BLR, yet within the DT at a distance of one parsec \citep{sahayanathan2012MNRAS,shah2019MNRAS,vittorini2017ApJ,dermer2014,yan2015MNRAS,roy2021MNRAS}.
\citet{shukla2020NatCo} reported the position of the $\gamma$-ray emission area to lie beyond the BLR by employing particle acceleration due to magnetic reconnection.
However, the flaring incident in June 2015 necessitates a highly condensed emission area situated within the BLR \citep{hayashida2015ApJ,pittori2018ApJ,ackermann2016ApJ,prince2020ApJ}. 
In addition, doubts have been raised regarding the particle acceleration mechanism through magnetic reconnection due to its low magnetization \citep{hu2020MNRAS,hu2021MNRAS,tolamatti2022APh}. Instead, the 2018 flare is being examined in light of the shock-in-jet model for particle acceleration \citep{tolamatti2022APh} or by an
injection of higher energy electrons \citep{wgg2022PASP}.\\

In the present work, we perform the long-term multi-wavelength
study of 3C 279 with the motivation of further understanding the
source behavior.
Epochs, where the weekly
averaged $\gamma$-ray flux is equal to or greater and than twice the 14 years
average $\gamma$-ray flux, are considered flaring/high activity states. The time periods with weekly
averaged $\gamma$-ray flux $\leq$  14 years
average $\gamma$-ray flux are chosen as quiescent/low activity states. Also, the epochs where the $\gamma$-ray flux is in between the flaring and quiescent were considered as intermediate states. Further, we identified those epochs where atleast the quasi-simultaneous data in optical/UV, X-ray and $\gamma$-ray data is available to construct the multi-band SEDs. Following this criteria, we found 168 such epochs.
Details of the multi-wavelength data analysis and results are given in Section 2. 
Our findings are discussed in Section 3 and the summary is provided in Section 4.\\

\section{Multiwavelength Observations and Data Reduction} \label{sec:obs_dat_red}
3C 279, is one among the brightest $\gamma$-ray blazars and was part target of interest sources at various continuous monitoring programs in different energy bands. The analysis of multi-wavelength data from different observatories are discussed below.

\subsection{\textit{Fermi}-LAT}
We used the data from {\it Fermi}-LAT \citep{atwood2009} covering the 
period from 4 August 2008 to 31 October 2022 ($\sim$14 years)
in the energy range of $100$ MeV to $300$ GeV.
We analyzed a $15^{\circ}$ region of interest (ROI)
centered on the source position, and $20^{\circ}$ source radius was used.
The \texttt{Fermitools} version $2.2.0$ software was used to carry out 
analysis of `pass $8$ P8R3' data \footnote{\url{https://fermi.gsfc.nasa.gov/cgi-bin/ssc/LAT/LATDataQuery.cgi}} (`diffuse class events; evclass = $128$, 
evtype = $3$'). The analysis was carried out using the \texttt{Fermipy} version 
$1.20$ software \footnote{\url{https://fermipy.readthedocs.io/en/latest/}} \citep{wood2017} with instrument response function (IRF) `P8R3\_SOURCE\_V3'. Good time intervals were obtained with the filter expression `DATA$\_$QUAL$>0$ \&\& LAT$\_$CONFIG$==1$'. To remove the contribution from 
the earth limb, a $90^{\circ}$ cut on the zenith angle was applied.  
Furthermore, a spatial binning of $0.1^{\circ}$ pixel$^{-1}$ and ten
logarithmically spaced energy bins per decade were chosen. 
The initial input model file was generated using `make4fglxml.py' (version: v01r09)\footnote{\url{https://fermi.gsfc.nasa.gov/ssc/data/analysis/user/}}, 
including all 4FGL-DR3 catalog sources \citep{Abdollahi2022ApJS} within $20^{\circ}$ of the ROI center.
Following the standard methodology, the ``Galactic diffuse emission model (\textit{gll\_iem\_v07.fits}) and extra-galactic isotropic diffuse emission (\textit{iso\_P8R3\_SOURCE\_V3\_v1.txt}) were included in the model file''.\\

The steps followed to calculate the $\sim$14 years average $\gamma$-ray 
flux of the source are discussed below. All sources' 
normalizations and spectral parameters within $15^{\circ}$ of ROI center were left to vary 
after the initial optimization using the `optimize' method of \texttt{Fermipy}. 
Furthermore, the Galactic and isotropic diffuse backgrounds' normalizations
and the spectral index of the Galactic diffuse background were left
free. The significance of source emissions is determined using test statistics defined as $TS = 2\times log(L_1/L_0)$, where $L_0$ and $L_1$ are the likelihoods of the model without source (null hypothesis) and the alternative likelihood (with source), respectively.
We froze all spectral
parameters, including normalization for sources with TS $<1$ 
and $N_{pred}$ value less than $10^{-3}$ counts.  
We generated a TS 
map and found additional sources in the ROI using the task
\texttt{find\_sources}. 
Eleven new sources (Table.~\ref{tab:new_gamma_srcs}) with $\sqrt{TS}\geq5$ were detected within the ROI and were added to the model. \\

\begin{table}[h]
\centering
\caption{Information on the new $\gamma$-ray sources detected during the $\gamma$-ray analysis.}
\begin{tabular}{ccc}\hline\hline
Source name      &RA (deg) &Dec (deg)   \\
\hline
PS J1307.1-0343 &$196.797$ & $-3.732$  \\
PS J1234.6-0430 &$188.675$ & $-4.514$  \\
PS J1304.7+0046 &$196.188$ & $0.769$ \\
PS J1245.9+0110 &$191.496$ & $1.171$  \\
PS J1302.3+0128 &$195.589$ & $1.468$  \\
PS J1238.4-1155 &$189.601$ & $-11.923$ \\
PS J1325.8-0405 &$201.457$ & $-4.090$  \\
PS J1322.7-0935 &$200.682$ & $-9.599$  \\
PS J1319.5-0042 &$199.887$ & $-0.708$ \\
PS J1317.9-1142 &$199.499$ & $-11.714$ \\
PS J1226.8-1311 &$186.711$ & $-13.194$ \\
\hline
\end{tabular}
\label{tab:new_gamma_srcs}
\end{table}

The target source was modeled using simple power law (PL: 
Equation ~(\ref{pl})) and log-parabola 
(LP: Equation ~(\ref{lp})) models:\\
\begin{equation}
\label{pl}
\frac{dF}{dE}=N \left(\frac{E}{E_{0}}\right)^{-\alpha}
\end{equation}
\begin{equation}
\label{lp}
\frac{dF}{dE}=N \left(\frac{E}{E_{b}}\right)^{-\alpha-\beta \log{\left(\frac{E}{E_{b}}\right)}} 
\end{equation}

Here, $\frac{\mathrm{d}F}{\mathrm{d}E}$ is the differential flux in 
ph cm$^{-2}$ s$^{-1}$ MeV$^{-1}$, $N$ is normalization factor in 
ph cm$^{-2}$ s$^{-1}$ MeV$^{-1}$, $E$ is the energy, 
$E_0$ and $E_{b}$ are the scale and break value, respectively, in the unit of MeV, $\alpha$ and
$\beta$ are the spectral parameters.\\

The 100 MeV to 300 GeV weekly averaged $\gamma$-ray light curve was created using the best fit model derived for the $\sim14$ years data set with 3C 279 modeled as a power-law spectra. We used the \texttt{Fermipy} package for the light curve creation. At each light curve bin, we left the spectral parameters and normalization of sources within $3^{\circ}$ of the ROI center as free; additionally, the normalization of all sources within $15^{\circ}$ of the ROI center, and Galactic and isotropic
emission’s normalization were left to vary. If the $TS\geq9$ (detection significance of $\sim3\sigma$), 
the target source was considered to be detected. 
Bins with $TS<9$ and/or $\Delta F_{\gamma}/F_{\gamma}>0.5$,
where $\Delta F_{\gamma}$ is the error estimate in the ﬂux $F_{\gamma}$ were not considered from the analysis.
For the various activity states, the $\gamma$-ray SEDs in 100 MeV$-$300 GeV were created in seven energy bands using the \texttt{sed} tool of \texttt{Fermipy}.

\subsection{Swift-XRT/UVOT}\label{sec:data_xray} 
We used data from the ``{\sl Swift} X-ray Telescope'' \citep[{\sl Swift}-XRT;][]{burrows2005}, which 
covers the energy range 
of 0.3$-$10 keV.
The online `{\sl Swift}-XRT data products generator'  was used to 
create the light curve with one bin per observation \citep{evans2007} and spectral data products \citep{2009MNRAS.397.1177E} 
for different activity states in the energy
range of 0.3--10 keV. 
This facility downloads data and chooses a suitable source and 
background regions depending upon the source count. To identify intervals 
affected by pile up, the tool first searches for times where the count rate 
within a 30 pixel radius circular region centered on the source is above 
0.6 counts s$^{-1}$ in photon counting (PC) mode or 150~counts s$^{-1}$ 
in window timing (WT) mode. After that, it accordingly chooses the source 
and background region of the annular shape in PC mode or a box annulus for 
WT mode \citep{2009MNRAS.397.1177E}. In our study, we used photon counting (PC) 
mode data. The source had only two WT mode observations. Furthermore, we used spectral data products, and the source 
spectra were binned to have $20$ counts per bin using \texttt{grppha}. 
Following \citet{paliya2021ApJS}, we also used the absorbed power-law model to 
fit the extracted spectrum using \texttt{XSpec} version: $12.13.0$ \citep{arnaud1996}. For the fitting, a Galactic absorption  value of $N_{H} = 2.24\times 10^{20}\, $cm$^{-2}$ 
\citep{HI4PICollaboration2016} was used. We used the `cflux' routine of 
\texttt{XSpec} to estimate the 0.3$-$10 keV source flux and photon index.\\

Complementing to {\sl Swift}-XRT, the simultaneous observations in the UV/optical bands from {\sl Swift}-UVOT \citep{roming2005} were also used in this work. We carried out the {\sl Swift}-UVOT data analysis using the \texttt{HEASoft} package
(v$6.31$), and $20211108$ version of \texttt{caldb}. Following \citet{bhattacharya2021}, ``\texttt{uvotimsum} 
task was used to merge the various observations during individual epochs considered for broadband SED construction. 
For photometry, a circular source region with 
$5$ arcsec radius centered at the source position and a background annular 
region with inner and outer radii of $15$ and $25$ arcsec were used.
\texttt{uvotmaghist} task was used
to create the light curve, and the \texttt{uvotsource} task was used to obtain the source magnitude.'' The magnitudes were converted to AB flux (erg cm$^{-2}$ s$^{-1}$) using the AB zero points taken
from \citet{breeveld2011}. The Galactic extinction was calculated using \citet{cardelli1989} and \citet{schlafly2011}. We note that all data Swift XRT/UVOT data from from blazar observations are also available through the Markarian Multiwavelength Data Center \citep{2024AJ....168..289S}.

\subsection{NuSTAR}
NuSTAR is a space-based X-ray telescope that uses grazing incidence optics to focus hard X-rays onto two focal plane modules, designated FPMA and FPMB \citep{2013ApJ...770..103H}. In the NuSTAR public archive, there are five observations of 3C 279 (ObsIDs 60701035002, 60701020001, 60601005001, 60002020004, and 60002020002); however, only three (ObsIDs 60701035002, 60002020002 and 60002020004) of them are in the period selected for theoretical modeling in our study.
The NuSTAR data are analyzed using the NuSTAR\_Spectra pipeline, which is designed to facilitate the systematic analysis of blazar observations from the NuSTAR public archive. NuSTAR\_Spectra retrieves data in the energy range of 3-79 keV from the NuSTAR public archive and processes it to produce high level scientific outputs. It calibrates and filters event files, detects sources, estimates background, and extracts spectra using both power-law and logarithmic parabola models. It calculates integrated fluxes in the 3-10 keV and 10-30 keV bands and generates SEDs for each observation. More details on NuSTAR\_Spectra are presented in \citet{2022MNRAS.514.3179M}. Moreover, all NuSTAR data from blazar observations are available through the Markarian Multiwavelength Data Center \citep{2024AJ....168..289S}.

\subsection{Optical/NIR Data}\label{sec:data_optical}
The data include both publicly available archival and WEBT observations, provided by our coauthor.  
3C 279 is one of the sources included in the WEBT\footnote{\url{http://www.oato.inaf.it/blazars/webt/}} target list. The WEBT data ($B$, $V$, $R$, $I$, $J$, $K$ and $H$) for this paper were published in \citet{boettcher2007}, \citet{larionov2008}, \citet{2010Natur.463..919A}, \citet{hayashida2012}, \citet{pittori2018ApJ}, and \citet{larionov2020}.
The source optical magnitude was derived with respect to Stars 1, 2, 3, and 5 calibrated by \citet{raiteri1998} in the $BVR$ bands, and by \citet{smith1998} in the $I$ band.
All datasets from the various observers participating to the WEBT projects are carefully assembled; offsets between datasets are corrected for; clear outliers are removed; data noise is reduced by binning data close in time from the same instrument. This processing is needed in order to get reliable light curves for an accurate data analysis. \\

We used optical/near infra-red (NIR) observations ($V$, $B$, $R$, $J$ and $K$ filters) from ``Small \& Moderate Aperture Research Telescope System'' \citep[SMARTS;][]{bonning2012}\footnote{\url{http://www.astro.yale.edu/smarts/glast/home.php}} 1.3 meter telescope.
In this work, observations in the $R$ and $V$ band filter from the ``Stewards Observatory program'' \citep{2009arXiv0912.3621S} were utilized.  
All the observations at Stewards Observatory are carried out using 
three telescopes, namely 2.3 m Bok telescope, 
1.54 m Kuiper telescope and 6.5 m MMT telescopes. We used user-provided photometric data from the Stewards Observatory\footnote{\url{http://james.as.arizona.edu/~psmith/Fermi/}}. 
The observations in the $V$ band were utilized from ``Catalina Real-Time Transient Survey'' 
\citep[CRTS;][]{drake2009}\footnote{\url{http://nesssi.cacr.caltech.edu/DataRelease/}}.
The optical data in $R$ band are obtained from ``Katzman Automatic Imaging Telescope''\citep[KAIT;][]{li2003}\footnote{\url{http://herculesii.astro.berkeley.edu/kait/agn/}}.
The data were corrected for Galactic extinction using \citet{schlafly2011}.

\subsection{Radio Data}\label{sec:radio_data}
The data used in this study include both publicly available archival and published observations, as well as data reduced from raw observations made at various radio telescopes. The archival data includes observations from: the Fermi-GST AGN Multi-frequency Monitoring Alliance (F-GAMMA), made at the Effelsberg telescope in Germany, with radio fluxes (4.8 and 21.7 GHz) published in \citet{angelakis2019}; the University of Michigan Radio Astronomy Observatory \footnote{\url{https://dept.astro.lsa.umich.edu/datasets/umrao.php}}
\citep[UMRAO;][]{aller1985} at the radio band 4.8 GHz; the 230 GHz data from the Submillimeter Array\footnote{\url{http://sma1.sma.hawaii.edu/callist/callist.html}} \citep[SMA;][]{gurwell2007} in Hawaii; and the 43 GHz data from the VLBA by the Boston University (BU) group: the VLBA–BU–BLAZAR programme\footnote{\url{http://www.bu.edu/blazars/BEAM-ME.html}}. Data analysis procedures were followed as explained in Section~2 of \citet{krishna2024MNRAS}.\\

In this work, we have utilized the relatively recent observations of 3C 279 made with Xinjiang Astronomical Observatory--Nanshan (XAO-NSRT), Urumqi, China (\citealt{sun2006,marchili2010}, and references therein). The flux density was measured in cross-scan mode, with each scan comprised of eight sub-scans (four in azimuth and four in elevation) over the source position. The data were acquired in this fashion at C-band (4.8 GHz) and K-band (23.6 GHz). The XAO-NSRT data used in this work were reduced by following the data calibration and reduction procedures as explained in Section~2 of \citet{marchili2011};  see also \citet{marchili2012} and \citet{liu2015}. Note that we have merged the observations from different observatories at 4.8 and 23 GHz in order to construct a complete light curve in each radio band. We used the VLBA data at 15 GHz\footnote{\url{https://www.cv.nrao.edu/MOJAVE/sourcepages/1253-055.shtml}}, which is part of the MOJAVE survey~\citep{lister2018ApJS}.\\

The 37 GHz observations were made with the 13.7 m diameter Aalto University Mets\"ahovi radio telescope, which is a radome enclosed Cassegrain type antenna in Finland  (60 d 13' 04'' N, 24 d 23' 35'' E). The measurements were made with a 1 GHz-band dual beam receiver centered at 36.8 GHz. The HEMPT (high electron mobility pseudomorphic transistor) front end operates at ambient temperature. The observations are Dicke switched ON--ON observations, alternating the source and the sky in each feed horn. A typical integration time to obtain one flux density data point is between 1200 and 1600 s. The detection limit of our telescope at 37 GHz is on the order of 0.2 Jy under optimal conditions. Data points with a signal-to-noise ratio $<4$ are handled as non-detections. The flux density scale is set by observations of the HII region DR21. Sources NGC 7027, 3C 274 and 3C 84 are used as secondary calibrators.  A detailed description of the data reduction and analysis is given in \citet{teraesranta1998}. The error estimate in
the flux density includes the contribution from the measurement rms and the uncertainty of the absolute calibration.


\section{Results and Discussions} \label{sec:result_discuss}
We have analyzed/included the available archival multiwavelength data of 3C 279 from 4 August 2008 to 31 October 2022 ($\sim$14 years) from {\sl Fermi}-LAT and Swift-XRT/UVOT, Nu-STAR, WEBT, Steward Observatory, SMARTS, KAIT, CRTS and radio data from XAO-NSRT, MOJAVE, VLBA-BU Blazar Monitoring Program, Aalto University Mets\"ahovi radio telescope, F-GAMMA, UMRAO, SMA. The data from all of the above telescopes were used to study the flux variation and to construct the multiband SEDs during different activity states of the source. A single-zone emission leptonic model was used to perform the multiwavelength SED modeling. The results are discussed in this section.

\subsection{Multiwavelength Variability}
\begin{figure*}
	\centering
	\includegraphics[scale=0.40]{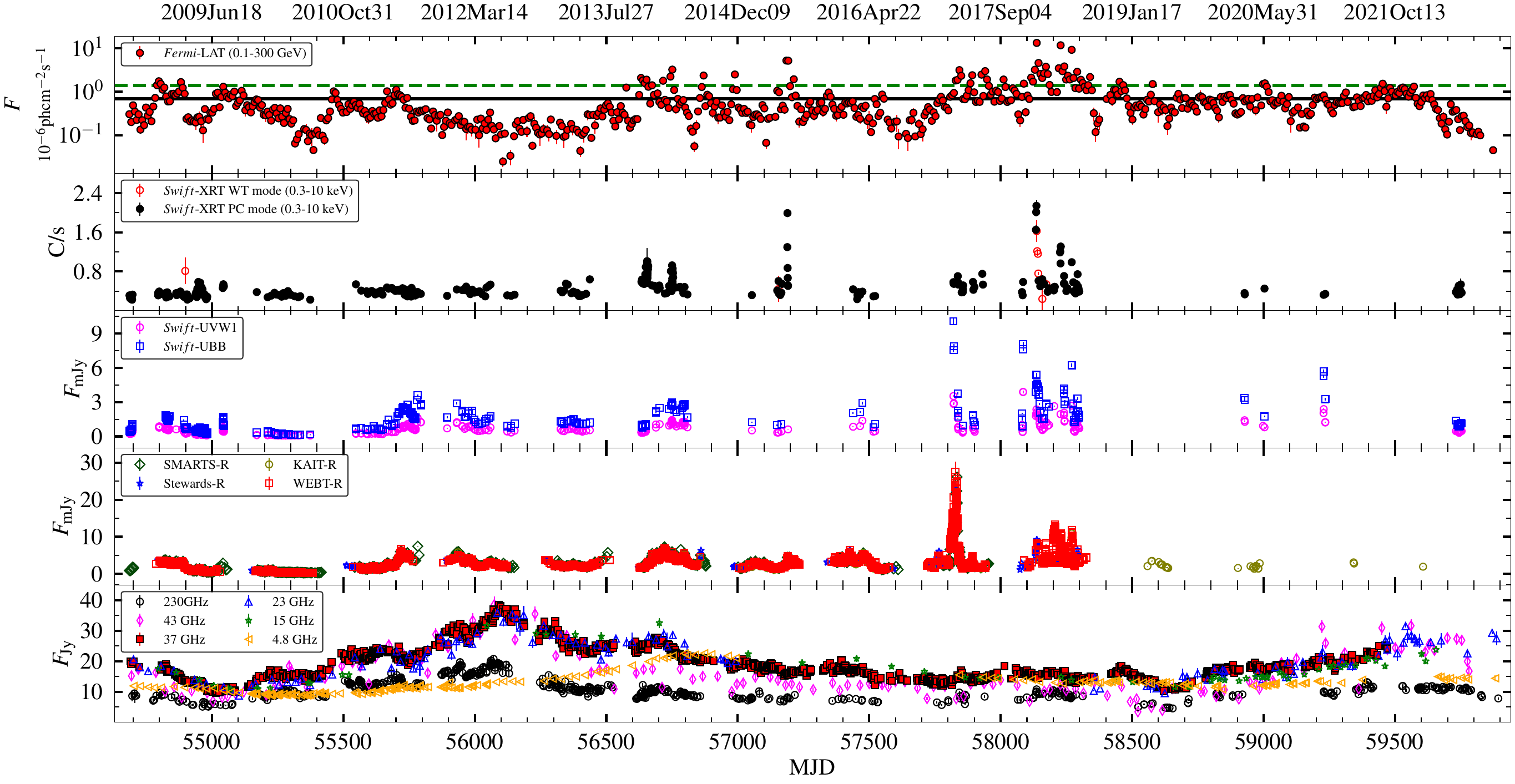}
	\caption{Multi-band light curve of 3C 279 between 4 August 2008 and 31 October 2022. 
    From the top to bottom: $\gamma$-ray light curve ($>100$ MeV) with the solid black and dashed green horizontal lines  
	representing the {\it Fermi}-LAT average $\gamma$-ray flux and twice the {\it Fermi}-LAT average $\gamma$-ray flux for $\sim14$ years of observations, 0.3-10 keV X-ray flux from {\sl Swift}-XRT, UV/optical flux from {\sl Swift}-UVOT, other optical/near-infrared data and multiband radio flux. }
    \label{fig1:multiplot_lc}
\end{figure*}

\begin{figure*}
	\centering
	\includegraphics[scale=0.50]{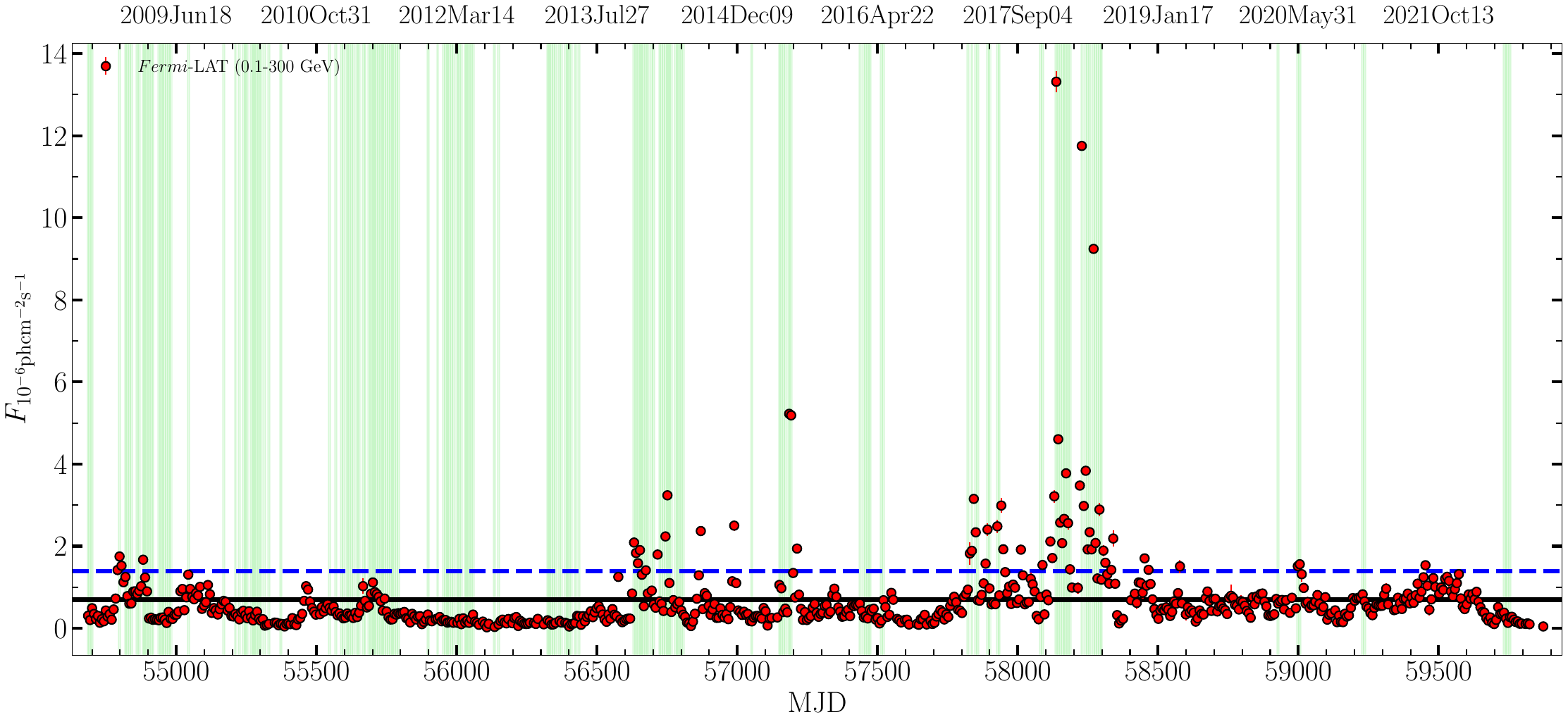}
	\caption{Weekly binned $\gamma$-ray light curve with the shaded regions representing the epochs with multiwavelength observations considered for the SED modeling. The horizontal
thick solid black and dashed blue lines indicate  average and twice the average $\gamma$-ray flux for $\sim 14$ years of observation, respectively.}
    \label{fig1a:gamma_weekly_lc}
\end{figure*}

In order to characterize the multiwavelength emission of the source, it is essential to produce its multiband flux variation with time. To achieve this, we created $\gamma$-ray and other wavelength light curves following the methods detailed in Section~\ref{sec:obs_dat_red}. 
The multiwavelength light curve of 3C 279 is shown in Figure~\ref{fig1:multiplot_lc}. Throughout the electromagnetic spectrum, the source exhibits flux variation and active flaring behavior. From the weekly binned $\gamma$-ray light curve (Figure~\ref{fig1a:gamma_weekly_lc}), epochs during which the weekly averaged $\gamma$-ray flux is equal to or exceeds twice the 14-year average $\gamma$-ray flux are classified as flaring or high activity states. The time periods with a weekly averaged $\gamma$-ray flux that is less than or equal to the 14-year average are designated as quiescent or low activity states. Additionally, epochs where the $\gamma$-ray flux falls between the flaring and quiescent states are categorized as intermediate states. Further, among these states, we consider those epochs for which at least quasi-simultaneous data in optical/UV, X-ray, and $\gamma$-ray wavelengths is available, enabling the construction of multiband spectral energy distributions (SEDs). Based on these criteria, we identified a total of 168 such epochs.\\

Following the above-mentioned definition of the different activity states in the $\gamma$-ray band, 
we found 108 quiescent/low, 27 intermediate, and 33 flaring/high activity states in the 168 time intervals. Furthermore, at each epoch, the entire energy range was divided into 7 logarithmically equal bands, and likelihood analysis was applied to produce the SED in the $\gamma$-ray band. 
At each of these epochs the source was modeled using PL, LP, and broken power-law (BPL), defined as:

\begin{equation}
\label{bpl}
\frac{dF}{dE}=N \times
\begin{cases}
     (E/E_{\rm break})^{\Gamma_{1}} & \text{if } E < E_{\rm break}   \\
     (E/E_{\rm break})^{\Gamma_{2}} & \text{otherwise} 
\end{cases}  
\end{equation}
\\
where $\frac{\mathrm{d}F}{\mathrm{d}E}$ is the differential flux in 
ph cm$^{-2}$ s$^{-1}$ MeV$^{-1}$, $N$ is normalization factor in 
ph cm$^{-2}$ s$^{-1}$ MeV$^{-1}$, $E$ is the energy, $\Gamma_{1}$ is index$1$, 
$\Gamma_{2}$ is index$2$ and $E_{\rm break}$ is the break value in the units of MeV.\\

\begin{figure*}
\begin{center}
	\includegraphics[scale=0.30]{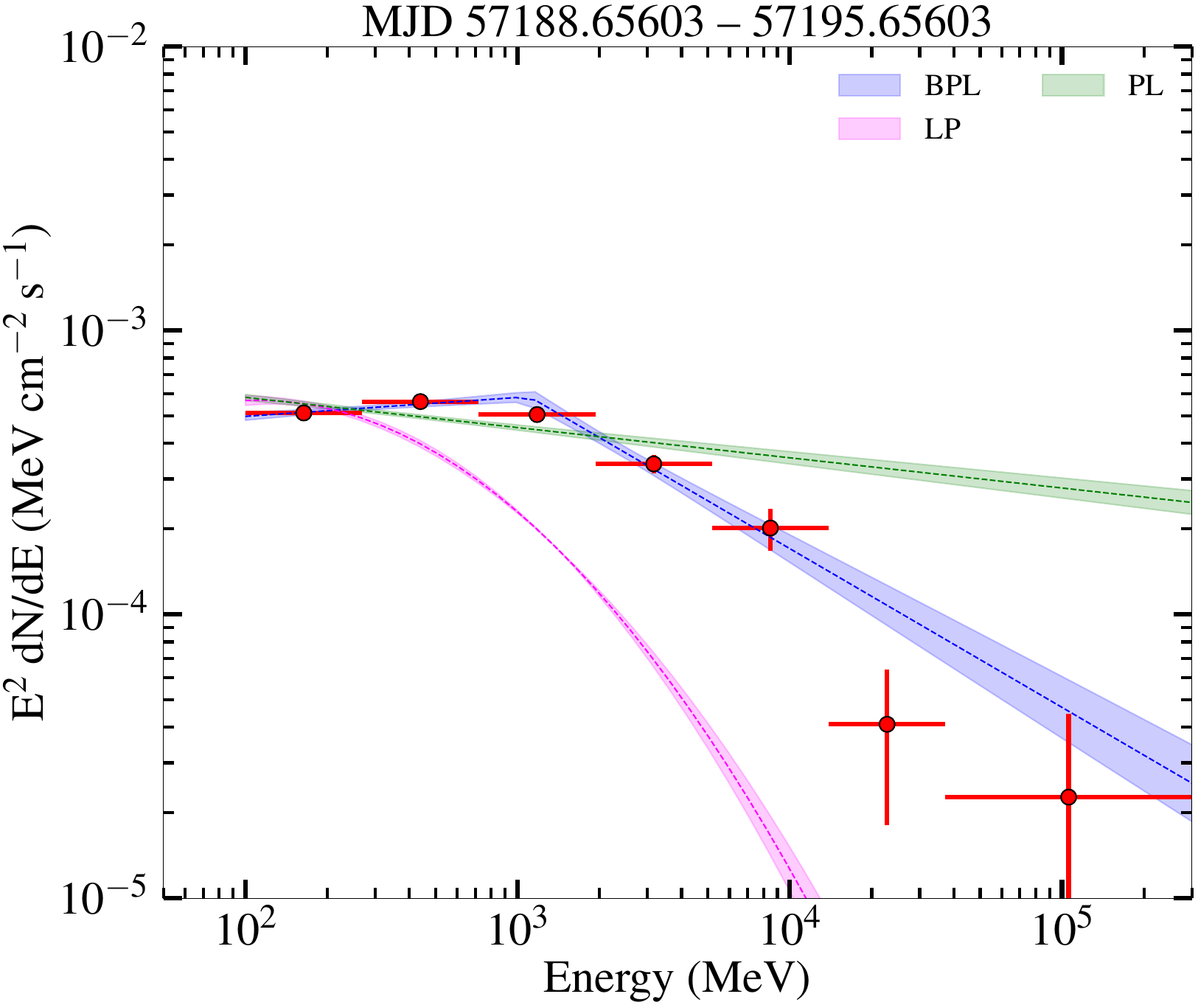}
        \includegraphics[scale=0.30]{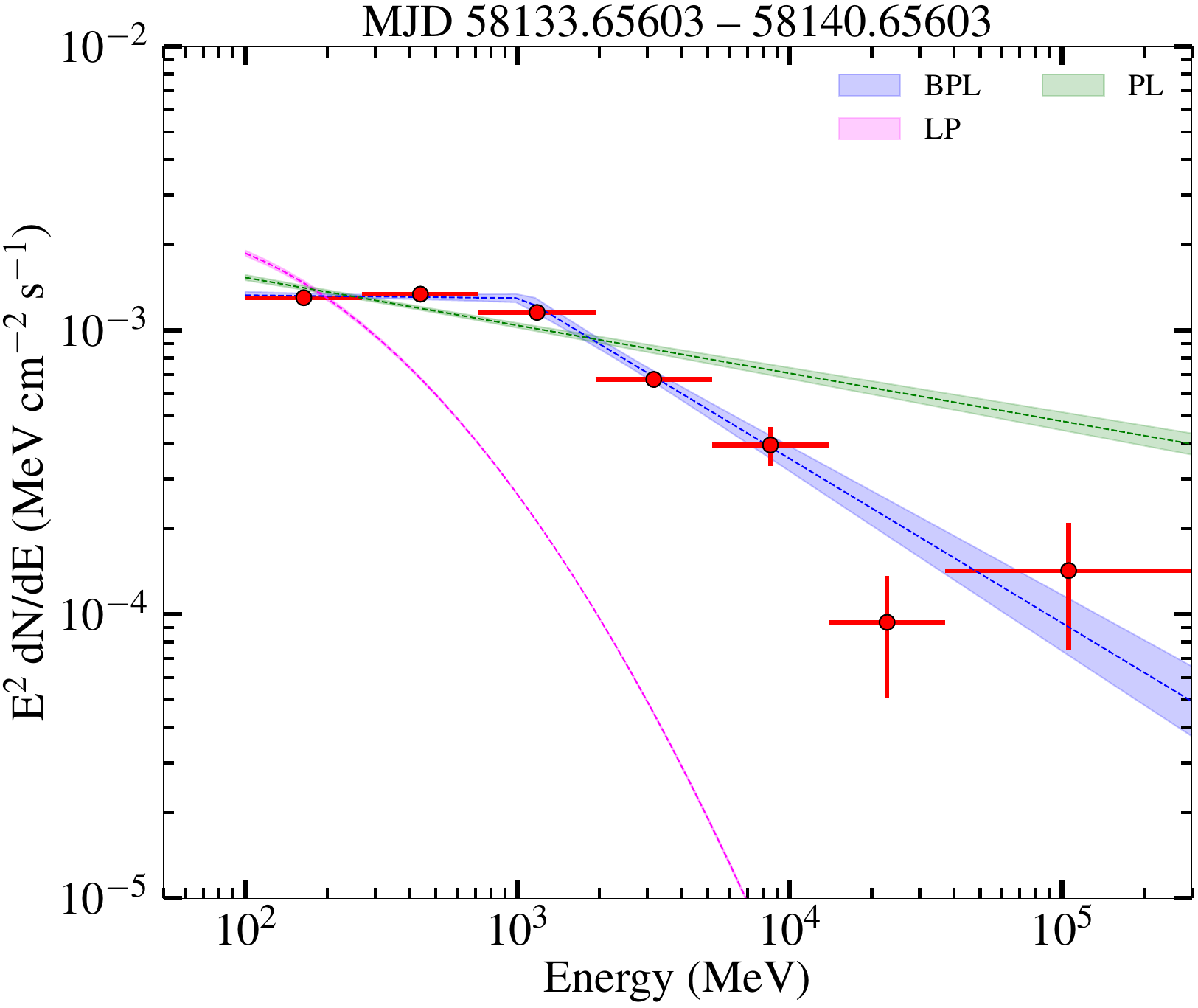}\\
        \includegraphics[scale=0.30]{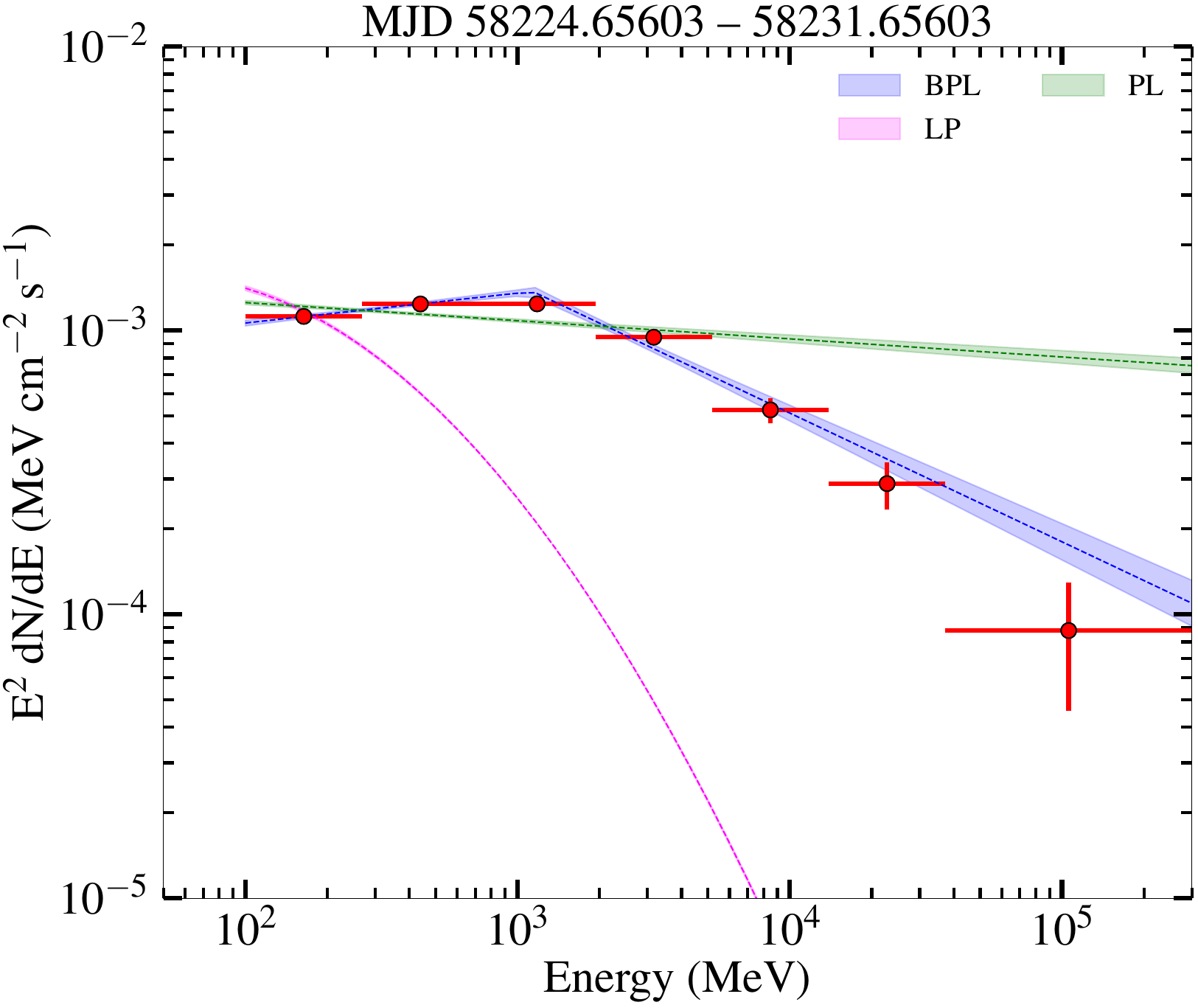}
        \includegraphics[scale=0.30]{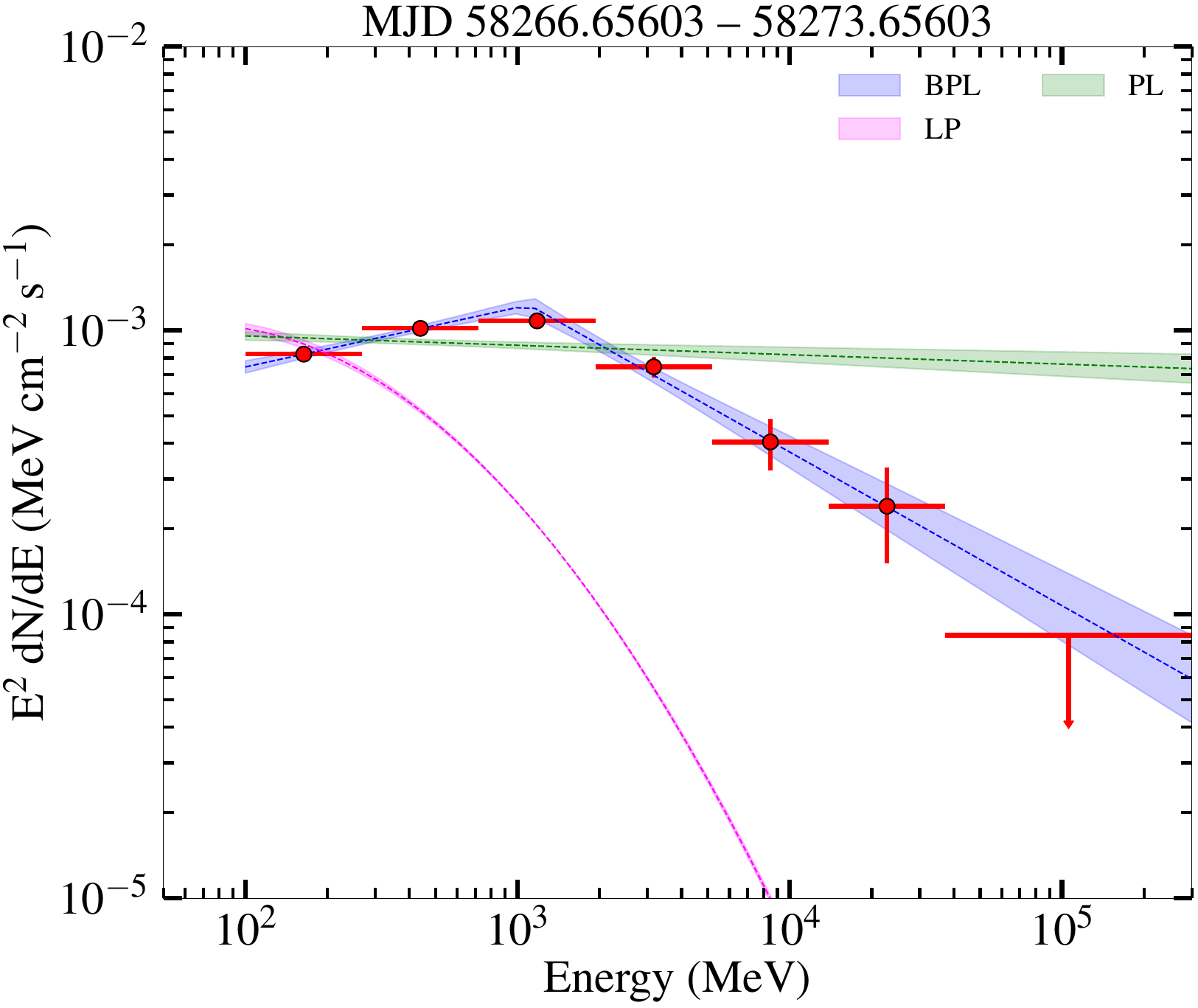}
\end{center}
	\caption{The $\gamma$-ray SEDs of 3C 279 in different epochs. The blue, magenta, green shaded area represents the BPL, LP and PL spectral fit, respectively.}
    \label{fig:gamma_ray_spectral_fit}
\end{figure*}

The 14 years average $\gamma$-ray spectra was found to be better represented by the LP model with flux of $(6.97\pm0.03)\times10^{-7}$ ph cm$^{-2}$ s$^{-1}$ and spectral parameter $\alpha=2.351\pm0.005$ and $\beta=0.071\pm0.002$. 
Using the PL model the average flux of $(7.20\pm0.03)\times10^{-7}$ ph cm$^{-2}$ s$^{-1}$ with $\gamma$-ray photon index $\alpha=2.277\pm0.003$ was estimated. 
The source was found to be significantly curved with $TS_{curve}=803.63$.
The results of the $\gamma$-ray spectral analysis for the 168 epochs are provided in the Table~\ref{tab:gamma_results}.
In the 168 epochs, the $\gamma$-ray spectra were well represented by the PL fit for 147 intervals, and 17 epochs were found to be significantly curved and better represented by BPL fit and four epochs with the LP fit. The curvature in the $\gamma$-ray spectrum can be asserted by calculating the TS$_{curve}$. Following \citet{nolan20122fgl} the TS$_{curve}$ can be defined as TS$_{curve}$ = 2$\times$($\log \mathcal{L}$~(LP/BPL) - $\log \mathcal{L}$~(PL)), where  $\mathcal{L}$ is the likelihood function.
The model parameters and the value of TS$_{curve}$ are provide in Table~\ref{tab:gamma_results}.
A model exhibiting a significantly high value of TS$_{curve}$ (with significance of the curvature $>5\sigma$) is regarded as the most suitable fit for the SED data points, indicating the presence of a spectral cut-off.
We noticed that during nearly all low and intermediate activity states the $\gamma$-ray spectra well fitted with the PL model. 
However, during the high activity state the spectra is well explained with PL and BPL model.\\ 

From the weekly binned $\gamma$-ray light curve (Figure~\ref{fig1a:gamma_weekly_lc}), we noticed that the source exhibits a very high flaring state on several occasions. These epochs include the flaring periods reported in previous studies \citep[e.g.,][]{hayashida2012,hayashida2015ApJ,paliya2015ApJa,paliya2016ApJ,paliya2015ApJb,ackermann2016ApJ,prince2020ApJ}. 
The highest $\gamma$-ray flux in the weekly binned light curve (Figure~\ref{fig1a:gamma_weekly_lc}) was observed during MJD 58133.65603 -- 58140.65603 with the flux of $127.40\pm2.00 \times 10^{-7}\:{\rm photon\:cm^{-2}\:s^{-1}}$ and the corresponding BPL spectral parameters were: $\Gamma_{1} = 2.01\pm 0.02$, $\Gamma_{2} = 2.58\pm 0.07$, $E_{\rm break} = 1.05\pm0.17$ GeV. The second highest $\gamma$-ray flux occurred during MJD 58224.65603 -- 58231.65603 with flux = $113.60\pm1.62 \times 10^{-7}\:{\rm photon\:cm^{-2}\:s^{-1}}$ and the BPL spectral parameters were: $\Gamma_{1} = 1.90\pm 0.02$, $\Gamma_{2} = 2.45\pm 0.04$, $E_{\rm break} = 1.14\pm0.11$ GeV. 
The third bright flare was observed during MJD 58266.65603 -- 58273.65603 with flux = $87.31\pm2.18 \times 10^{-7}\:{\rm photon\:cm^{-2}\:s^{-1}}$, $\Gamma_{1} = 1.79\pm 0.04$, $\Gamma_{2} = 2.54\pm 0.09$, $E_{\rm break} = 1.10\pm0.16$ GeV. Also, the BPL model was found to be better represent the $\gamma$-ray spectra during the period MJD 57188.65603 -- 57195.65603 (flare during June 2015) with flux = $51.13\pm0.88\times 10^{-7}\:{\rm photon\:cm^{-2}\:s^{-1}}$, $\Gamma_{1} = 1.93\pm 0.03$, $\Gamma_{2} = 2.56\pm 0.07$, $E_{\rm break} = 1.10\pm0.16$ GeV. The different $\gamma$-ray spectral model fit to the SED for the above mentioned periods are shown in Figure~\ref{fig:gamma_ray_spectral_fit}. Similarly, during the other 13 epochs, BPL spectral fit better represented the $\gamma$-ray spectrum (Table~\ref{tab:gamma_results}).\\

The existence or lack of curvature in the $\gamma$-ray SEDs is essential for determining the position of the emission region. A break in the $\gamma$-ray spectrum is anticipated when the emission source resides within the broad line region (BLR), due to photon-photon pair production \citep{liu2006ApJ,Poutanen2010ApJ,stern2011MNRAS, 2020MNRAS.496.5518S}. For photons with energy $>20$ GeV, the BLR region is opaque; therefore, a curvature or break in the $\gamma$-ray spectrum can be observed above 20 GeV \citep{liu2006ApJ}. In our study, we did not observe any spectral break above 20 GeV and moreover, except for a few cases, the $E_{\rm break}$ value was between $1-2$ GeV, almost constant irrespective
of the different high activity states. Such findings are consistent with previous studies reported on this source by \citet{paliya2015ApJb,prince2020ApJ} and 3C 454.3 by \citealt{sahakyan2021MNRAS}, and references therein. An alternative way to explain such curvature or the cut-off is to say that if a cut-off is already present in the particle distribution under consideration \citep{prince2020ApJ}.\\

\begin{figure*}
	\centering
	\includegraphics[scale=0.3]{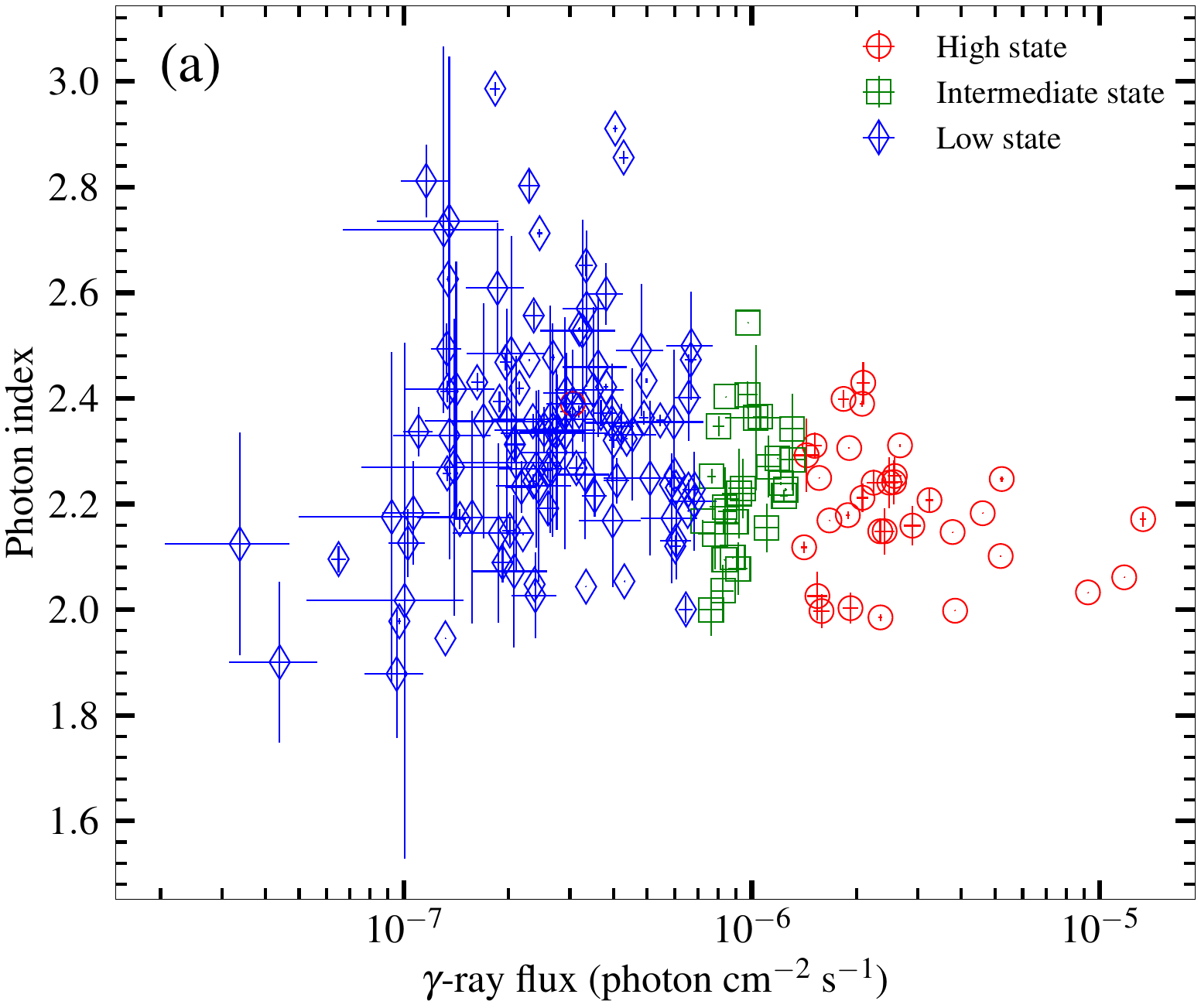}
        \includegraphics[scale=0.3]{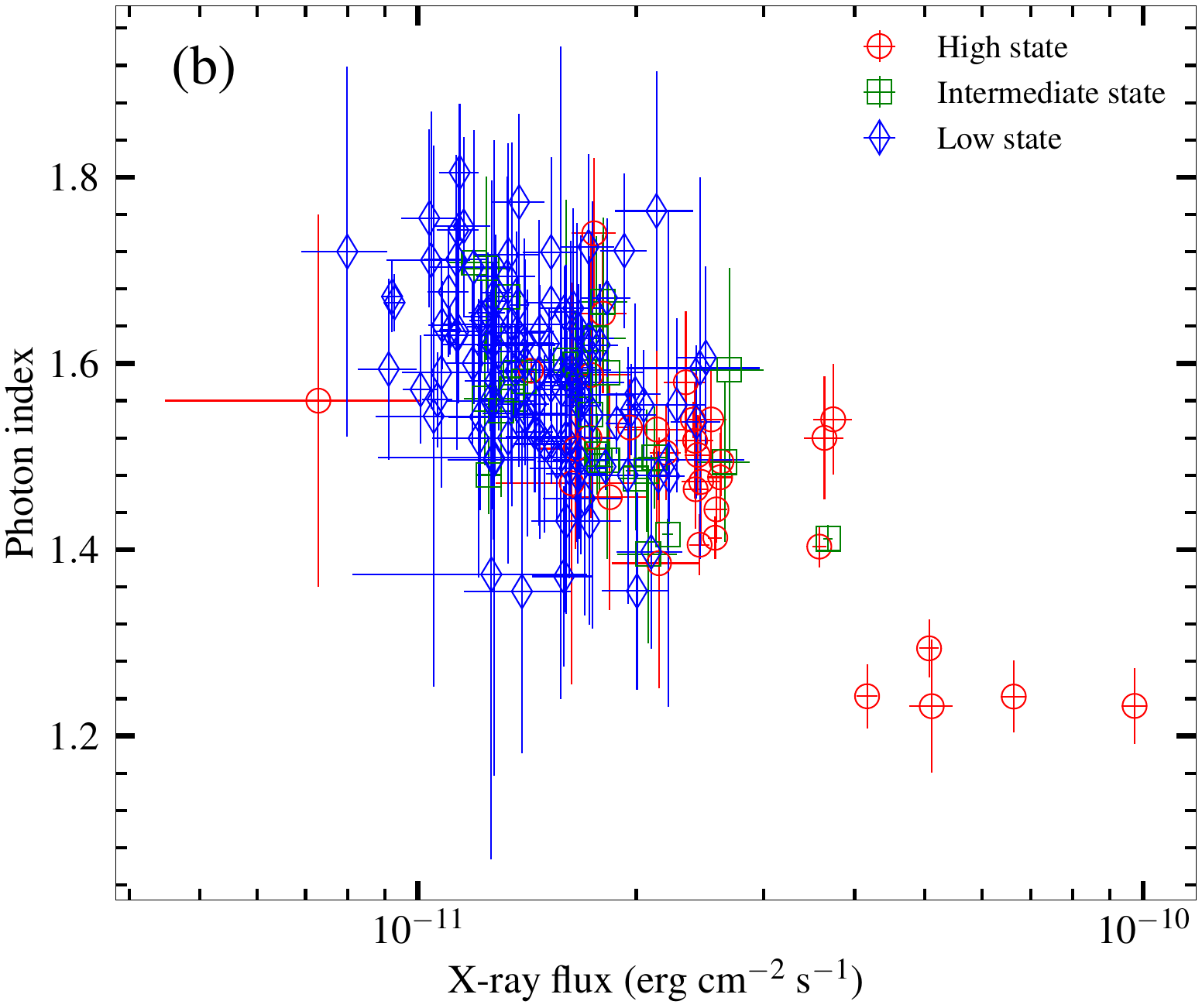}
	\caption{Photon index versus flux during different activity states estimated in weekly bins. (a) the $\gamma$-ray photon index versus the flux. (b) the X-ray photon index versus the flux.}
    \label{fig:flux_index_plot}
\end{figure*}

The $\gamma$-ray photon index versus flux estimated during the different activity states is
presented in Figure~\ref{fig:flux_index_plot} (top panel) and linear-Pearson correlation test was used to check the possible presence of correlation. The test yielded $r_{p}=-0.24$, indicating a weak negative correlation between the flux and photon index, i.e., the flux increases with decreasing (hardening) photon index. 
In the X-ray band (Figure~\ref{fig:flux_index_plot} lower panel), 3C 279 shows the typical FSRQ behavior with a hard X-ray photon index of $\sim1.6$. The lowest value of photon index $1.23$ is noticed during the high activity states with X-ray flux reaching upto $9.73 \times 10^{-11}\:{\rm erg\:cm^{-2}\:s^{-1}}$. The linear-Pearson correlation test yielded $r_{p}=-0.61$, showing negative correlation pattern.\\

\begin{figure*}
	\centering
	\includegraphics[scale=0.40]{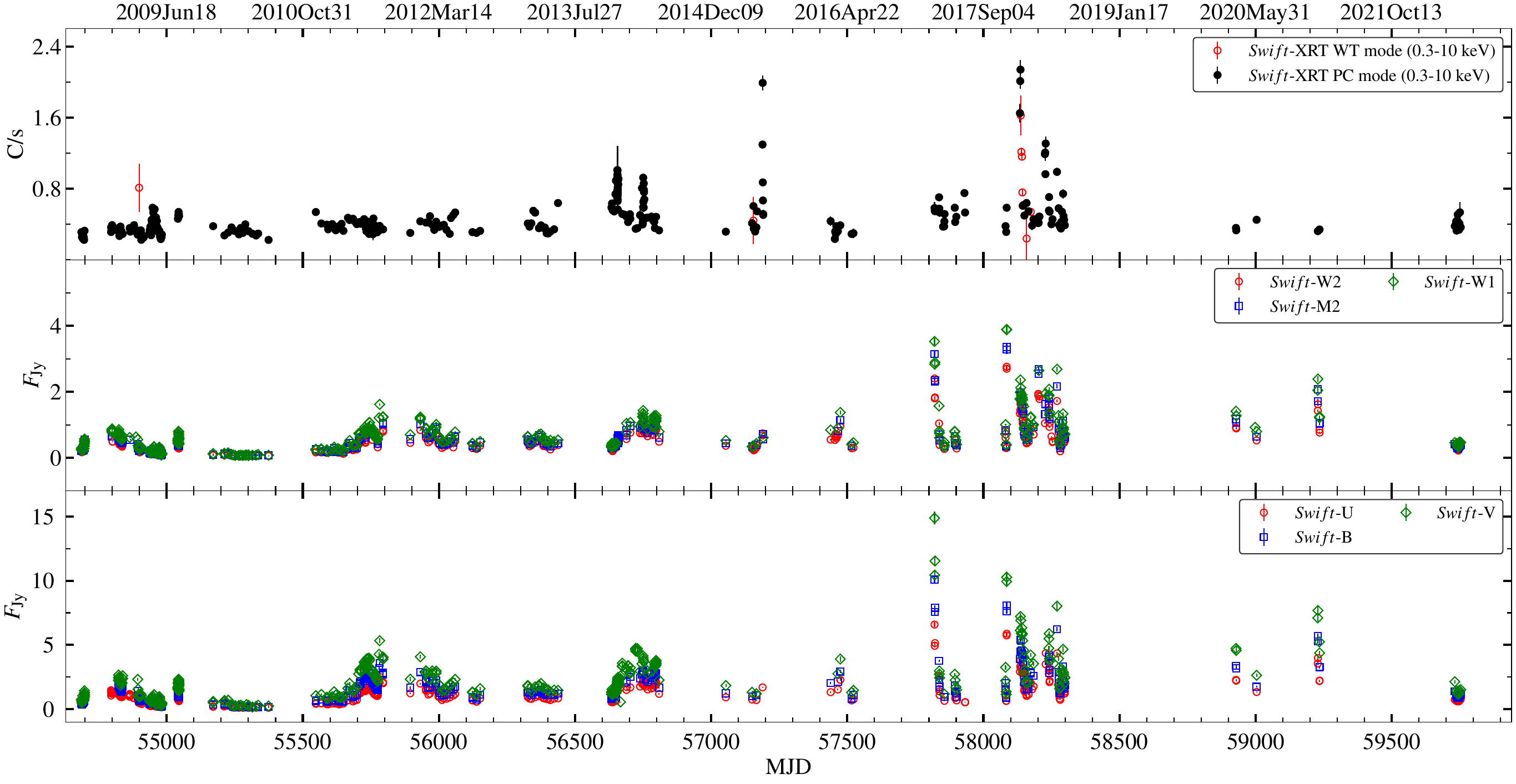}
	\caption{Multiband light curve of 3C 279 from {\sl Swift}-observations. From the top to bottom: 0.3-10 keV X-ray flux, UV flux and optical flux, respectively.}
    \label{fig2:multiband_plot_all_swift_data}
\end{figure*}

\begin{figure*}
	\centering
	\includegraphics[scale=0.40]{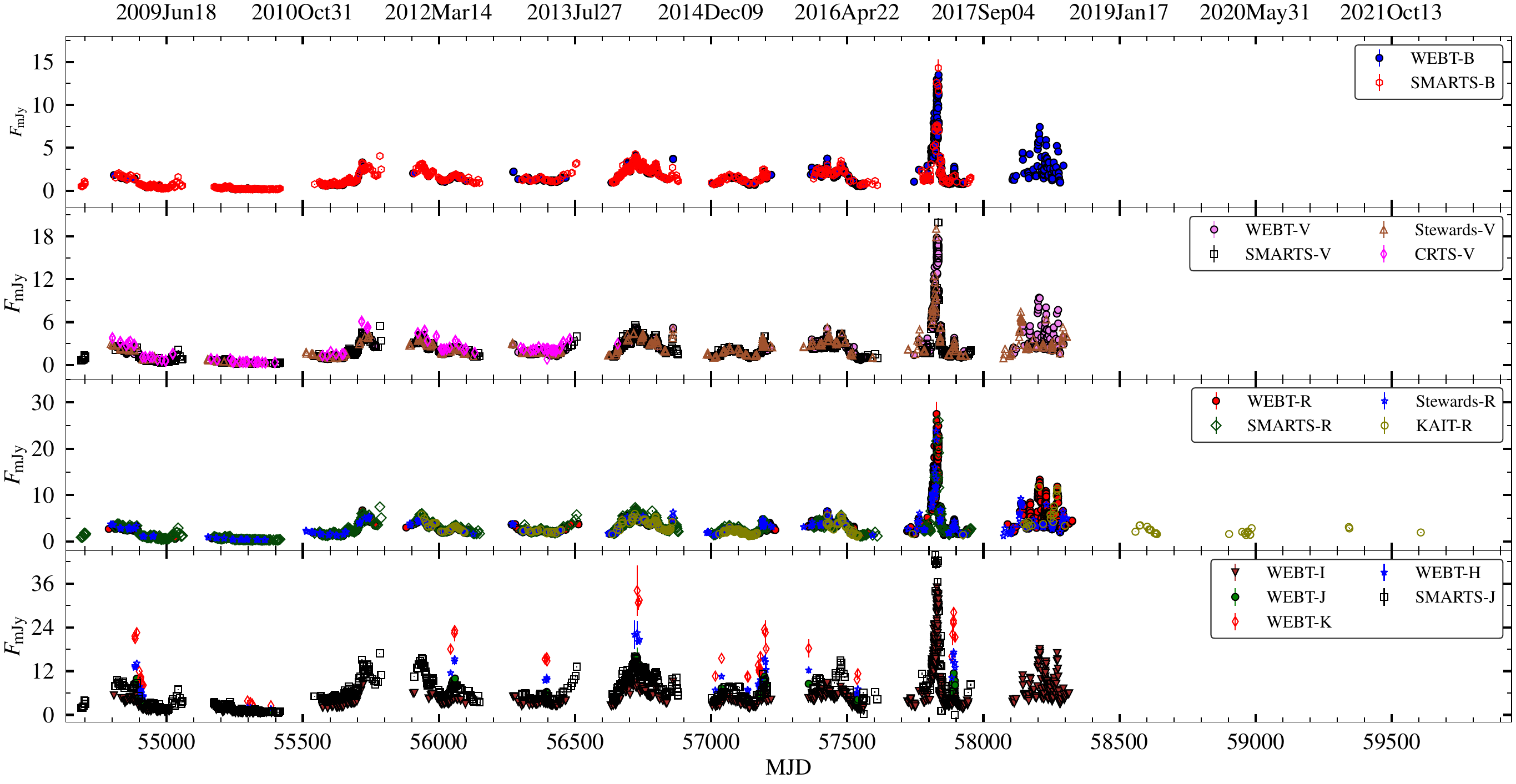}
	\caption{Optical and near-infrared light curves of 3C 279 from the WEBT campaign, SMARTS, KAIT and CRTS observatories.}
    \label{fig3:multiband_plot_all_optical_data}
\end{figure*}

\begin{figure*}
	\centering
	\includegraphics[scale=0.40]{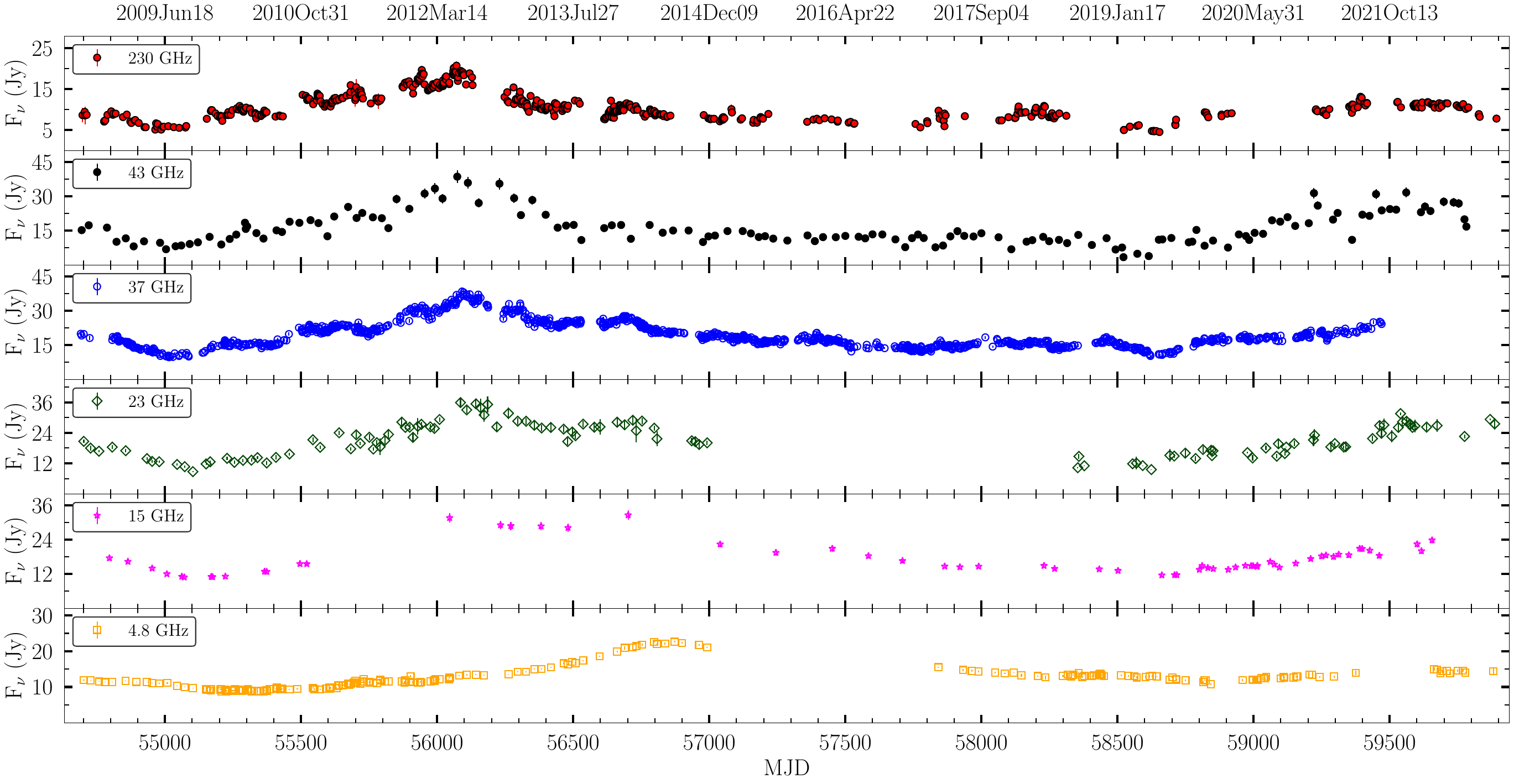}
	\caption{Radio light curves of 3C 279 during the time interval 2008–2022.}
    \label{fig5:multiband_plot_all_radio_data}
\end{figure*}

The detailed UV/optical, near-infrared, and radio band light curves during 2008-2022 are shown in Figure~\ref{fig2:multiband_plot_all_swift_data}, Figure~\ref{fig3:multiband_plot_all_optical_data} and Figure~\ref{fig5:multiband_plot_all_radio_data}, respectively. From the visual inspection, it is evident that the source is highly variable and shows prominent flare-like features in UV/optical and near-infrared compared to the radio band. However, the light curve's fluctuation results from both statistical uncertainty in the flux measurements and variation in the intrinsic physical processes. To identify the presence of true intrinsic variability in the light curve, we have used the fractional root mean square variability parameter (F$_{var}$), since it considers the uncertainties in flux measurement. Here we use the relation given in \citet{vaughan2003MNRAS}, which is defined as:
\begin{equation} \label{Eq:frac_var}
 F_{var} = \sqrt{\frac{S^2 - \sigma^2}{r^2}} \\
\end{equation}

\begin{equation}
 err(F_{var}) = \sqrt{  \Big(\sqrt{\frac{1}{2N}}. \frac{\sigma^2}{r^2F_{var}} \Big)^2 + \Big( \sqrt{\frac{\sigma^2}{N}}. \frac{1}{r} \Big)^2     } \\
\end{equation}

where, $\sigma^2_{XS}$ = S$^{2}$ -- $\sigma^2$, is called excess variance, S$^{2}$ is the sample variance, $\sigma^2$ is the mean square uncertainties
of each observation and $r$ is the sample mean. The estimated fractional variability values for different light curves from $\gamma$-ray to radio wavelength are given in Table~\ref{table:frac_var}.\\

\begin{table}
\centering
\caption{Fractional variability in different wavebands estimated for the time interval 2008-2022.}
\begin{tabular}{c c c }\hline\hline
 Waveband  & F$_{var}$ & err(F$_{var}$) \\
 \hline
 $\gamma$-ray & 1.422  & 0.004  \\ 
 X-ray & 0.518 & 0.005 \\
 {\sl Swift}-U &0.814  & 0.002 \\
 {\sl Swift}-B &0.875 & 0.002 \\
 WEBT-B   & 0.949     & 0.002   \\
 SMARTS-B   & 0.870    & 0.002    \\
 {\sl Swift}-V &0.781  & 0.003\\
 WEBT-V   & 0.963     & 0.001  \\
 SMARTS-V & 0.874    & 0.001    \\
 Stewards-V  &0.747     &0.001  \\
 CRTS-V  & 0.647       &0.004   \\
 WEBT-R   & 0.913    &  0.001   \\
 SMARTS-R  & 0.836    & 0.002    \\
 Stewards-R  & 0.739    & 0.001    \\
 KAIT-R  &0.535    &0.001     \\
 WEBT-I  & 0.958    &  0.001  \\
 WEBT-J  & 0.536    & 0.011    \\
 WEBT-K  & 0.514      & 0.015    \\
 WEBT-H  & 0.518    & 0.014    \\
 SMARTS-J & 0.921    & 0.001   \\
 {\sl Swift}-W1 &0.840 &0.002  \\
 {\sl Swift}-M2 &0.872 &0.003\\
 {\sl Swift}-W2 &0.912 & 0.002 \\
 230 GHz & 0.294  & 0.002 \\
 43 GHz & 0.436 & 0.006 \\
 37 GHz & 0.283  & 0.001 \\
 23 GHz & 0.295 &0.007   \\
 15 GHz & 0.303 &0.007  \\
 4.8 GHz & 0.245 & 0.001 \\
 \hline
\end{tabular}
\label{table:frac_var}
\end{table}

The source variability in the $\gamma$ -ray band is observed to be greater than $100\%$, followed by X-ray at more than
$50\%$. In UV and optical the source variability varied between $\sim50\%$ to $\sim90\%$, including the near-infrared band. In the radio band, the fractional variability ranges from $\sim20\%$ to $\sim30\%$. The general trend reveals that the $F_{var}$ increases with the increasing energy band, indicating a higher number of particles are injected into the jet, resulting in high energy emission. This pattern was reported for this source in the earlier study by \citet{prince2020ApJ} with multiband observations covered between 2017 and 2018. However, this may not always be the case; for example decrease in fractional variability with higher frequency band (NIR, optical, and UV) was reported by \citet{bonning2009ApJ} for the source 3C 454.3, suggesting the presence of constant thermal emission from the accretion disk.

\subsection{Modeling the Multiwavelength SEDs}
SEDs during all the different epochs were fitted with one-zone leptonic synchrotron and external inverse Compton model. 
We used the publicly available source package \textit{Jetset} version 1.2.2 to model the broadband
SED \citep[][for the alternative model that uses Convolutional Neural Networks see \citealt{2024ApJ...963...71B} and \citealt{2024ApJ...971...70S}]{tramacere2009,tramacere2011,tramacere2020}.
The contributions from synchrotron photons produced inside the jet \citep[synchrotron self-Compton or SSC;][]{jones1974ApJ,marscher1985}, external photon field coming directly from the accretion disk \citep{ghisellini2010}, and reprocessed disk photon field coming from BLR \citep{sikora1994} and dusty torus \citep[external Compton or EC;][]{sikora2002ApJ} are all considered in the inverse Compton process. 
A spherical plasma blob with radius \lq $R$' at a distance \lq $d$' from the central supermassive black hole of mass \lq $M_{BH}$' is assumed to be the source of broadband emission in this model.
The blob has a bulk Lorentz factor \lq $\Gamma$' and is moving relativistically along the jet.
It is considered that the magnetic field ($B$) within the blob is same and isotropic throughout. 
A non-thermal population of electrons with an energy distribution of broken power-law form was considered.
\begin{equation} \label{eq:3}
n\left(\gamma \right) = \left\{ \begin{array}{cc} 
k \gamma^{-p} & \hspace{5mm} \gamma \leq \gamma_{b} \\
k \gamma_{b}^{(p_1-p)} \gamma^{-p_1} & \hspace{5mm} \gamma > \gamma_{b} \\
\end{array} \right.
\end{equation}

The low-energy hump in the SED is caused by the cooling of the electron population through synchrotron emission, resulting from interactions with the magnetic field inside the emission blob. Synchrotron photons interact with relativistic electrons, leading to their Compton upscattering (SSC).
Photons originating from the accretion disk can either enter the emission region directly or be reprocessed by the BLR and dusty torus before reaching the emission blob, where they contribute to Compton upscattering (EC).
As a result, both SSC and EC processes generate the high-energy peak in the SED.
The radiation from the emission blob is Doppler boosted in observer's frame by a factor $\delta$ as it moves, along our line of sight:
$\delta = [\Gamma\left(1-\beta cos\theta\right)]^{-1}$,
where $\Gamma$ is the bulk Lorentz factor of the emitting blob and $\theta$ is the angle between our line of sight and the the jet axis. The value of $\theta (\text{degree})=2.4$ was taken from \citet{hovatta2009}. In our study, we assumed one proton per 10 electrons inside emitting blob \citep{ghisellini2014Natur}.\\

In \citet{ghisellini2009}, expressions for the distances of the broad-line region (BLR), \( R_{\text{BLR}} \), and the dusty torus, \( R_{\text{DT}} \), from the central engine were derived as functions of the accretion disk luminosity, \( L_d \). These relations are given as follows:
\begin{equation}
R_{BLR} = 10^{17} \times (L_d/10^{45})^{1/2}
\end{equation}
\begin{equation}
R_{DT} = 2.5 \times 10^{18} \times (L_d/10^{45})^{1/2}
\end{equation}
Here it is assumed that BLR is a thin spherical shell of ionized gases, the inner and outer radii of BLR was chosen to be around $R_{BLR}$ such that, $R_{BLR}$ = ($R_{BLR}^{in}$+$R_{BLR}^{out}$)/2 and ($R_{BLR}^{out}$-$R_{BLR}^{in}$) = $2\times10^{16}$ cm. A typical temperature of 1000 K was used for the dusty torus \citep{roy2021MNRAS}.
The accretion disk model was a multi-temperature blackbody, where the temperature of any part of the disk was determined by its distance from the core as,
\begin{equation}
T^4\left(r \right) = \frac{3 R_S L_d}{16 \epsilon \pi \sigma_{SB} r^3}\left(1 - \sqrt{\frac{3 R_S}{r}}\right)
\end{equation}
Here, $\epsilon$ represents the accretion efficiency, fixed at 0.08 \citep{roy2021MNRAS}, $R_S$ is the 
Schwarzschild radius, and $\sigma_{SB}$ stands for the Stefan-Boltzmann constant. The model assumes the accretion disk spanned a region from $3 R_S$ to $500 R_S$ away from the central engine. 
The reprocessing of the accretion disk luminosity by the BLR ($\tau_{BLR}$) is considered to be a blackbody that peaks at the rest-frame frequency of Lyman-$\alpha$ \citep{ghisellini2009}.
The energy density observed in the comoving frame within \( R_{\rm BLR} \) with distance of the emission region from the central black hole, along
the jet axis ($d$) can be approximated as follows \citep{ghisellini1996MNRAS}:
\begin{equation}
U^\prime_{\rm BLR} \, \sim \,  { 17 \Gamma^2 \over 12} \, { \tau_{\rm BLR} L_{\rm d}
\over 4\pi R_{\rm BLR}^2 c} 
\,\,  \quad d < R_{\rm BLR}
\label{ublr1}  
\end{equation}
The emission profile of dusty torus is considered as a simple blackbody peaking at temperature $T_{torus}$. The fraction of $L_{d}$ intercepted by the torus and re-emitted in IR is represented as 
$\tau_{\rm DT}$. The comoving radiation energy density of torus component scales as  $U^\prime_{\rm BLR}$, but substituting \( R_{\rm BLR} \) with  \( R_{\rm DT} \). 
\begin{equation}
U^\prime_{\rm IR} \, \sim  \, { \tau_{\rm DT} L_{\rm d}\, \Gamma^2 
\over 4\pi R_{\rm DT}^2 c} 
\,\, \quad d < R_{\rm DT}
\end{equation}
For emission region at distances greater than \( R_{\rm BLR} \) or \( R_{\rm DT} \), energy density the relations are followed as provided in \citep{ghisellini2009}. 
Doppler boosting by a factor of approximately $\Gamma^2$ in the blob's comoving frame enhances photon fields from the BLR and dusty torus while the blob is within these structures.
The energy densities and luminosities in the blob comoving frame are initially computed by this model.
After that, luminosities are converted into flux in the observer frame by estimating luminosity distance ($D_L$) from the given redshift using the cosmological model with $\Omega_\Lambda = 0.685$, $\Omega_M = 0.315$ and Hubble's constant H$_0$= 67.3 km s$^{-1}$ Mpc$^{-1}$ \citep{planck2014}.

\subsection{SED Modeling Approach}
We adopt the values of lower boundary of $R_{BLR}^{in} = 0.9\times10^{17}$ cm and $R_{BLR}^{out} = 1.1\times10^{17}$ cm and which reflects 10 percent ($\tau_{BLR}$ = 0.1) of the accretion disk luminosity $L_{d}$ (10$^{46}$ erg s$^{-1}$) = 0.1 \citep{roy2021MNRAS}. The disk emission is approximated as a mono-temperature blackbody. To reduce the number of free parameters, it is assumed that the emission region size is $R=6.0\times10^{16}$ cm \citep{roy2021MNRAS} , which corresponds to hour scale variability as observed in the $\gamma$-ray band. As an initial set-up of the model, the parameters estimated in the quiescent state by \citet{roy2021MNRAS} were used. The remaining free parameters, namely the
electron particle number density ($N_{e}$), $p_{1}$, $p_{2}$, $\gamma_{min}$, $\gamma_{b}$, $\Gamma$, $B$, $\gamma_{max}$ were varied to match the broadband SED in each epoch. 
The distance of the emitting blob from the central black hole was kept constant at $d$ (10$^{17}$ cm) = 15.0, which was found to be consistent for all epochs, except for two of the very bright $\gamma$-ray states. The other fixed model parameters used for fitting the SEDs during all epochs are: M$_{BH}$  (M$_{\sun}$) = $7.9\times10^{8}$ \citep{Gu2001,roy2021MNRAS}, $R_{DT}$ = $2.5\times10^{18}$ cm and $\tau_{DT}$ = 0.4. 
It should be noted that the radio data are not used in the model fitting. Emission at low frequencies is affected by synchrotron self-absorption and so most likely it is produced in a more extended region \citep{sahakyan2021MNRAS, roy2021MNRAS}.\\

\subsection{SED Fit Results and Origin of Multiwavelength Emission}

\begin{figure*}[h]
	\centering
\includegraphics[width=0.41\textwidth,angle=0]{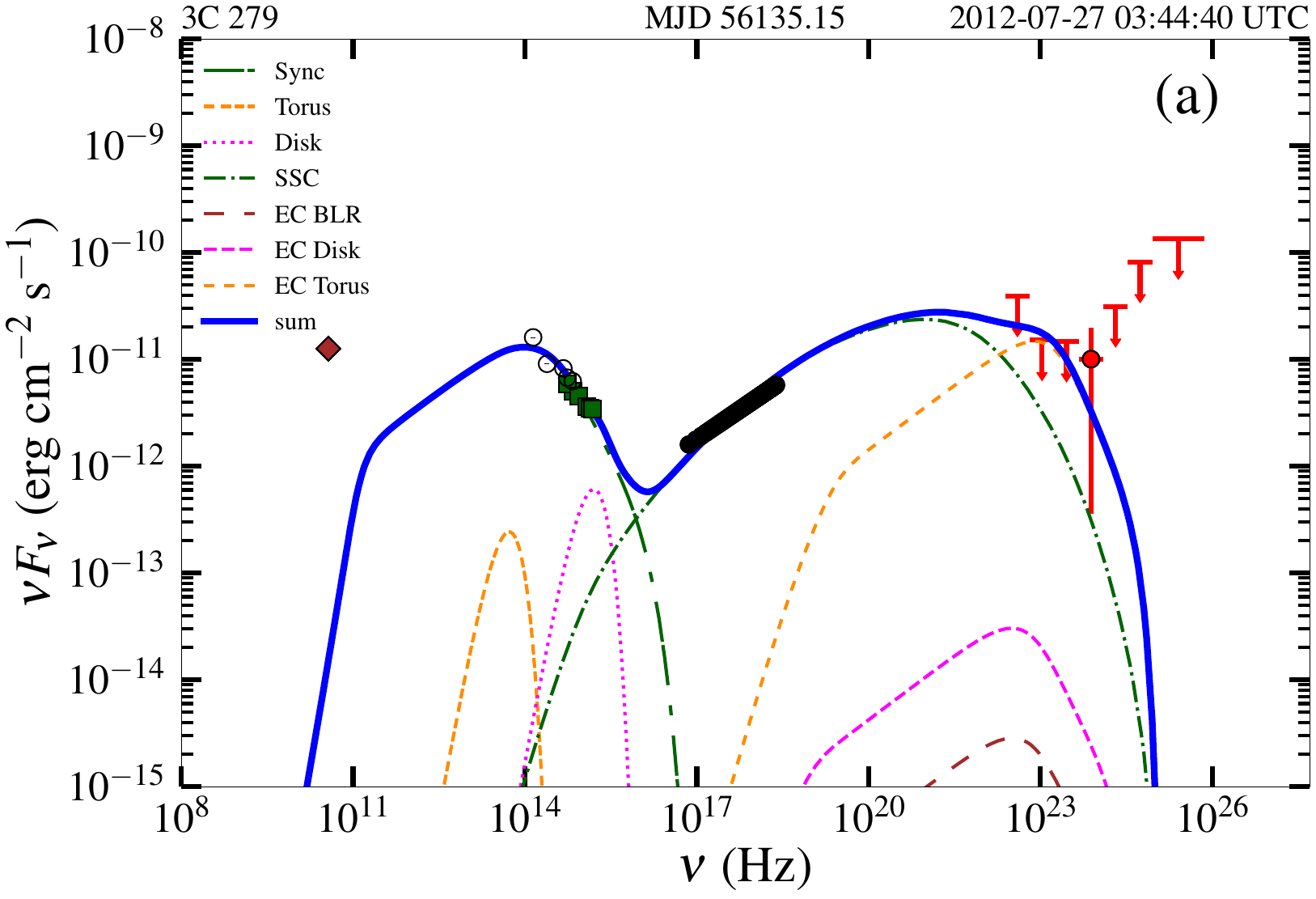}
\includegraphics[width=0.41\textwidth,angle=0]{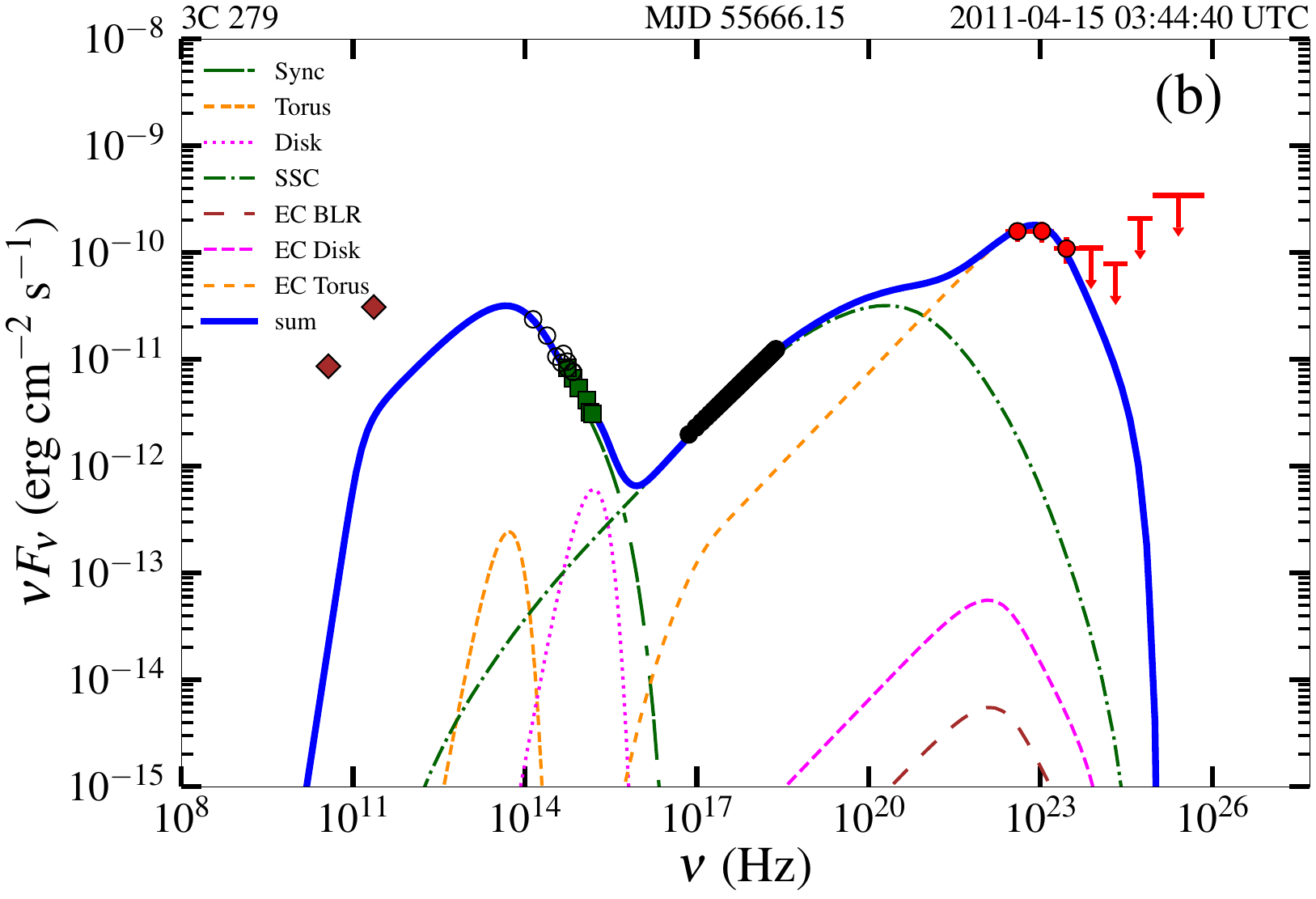}\\
\includegraphics[width=0.41\textwidth,angle=0]{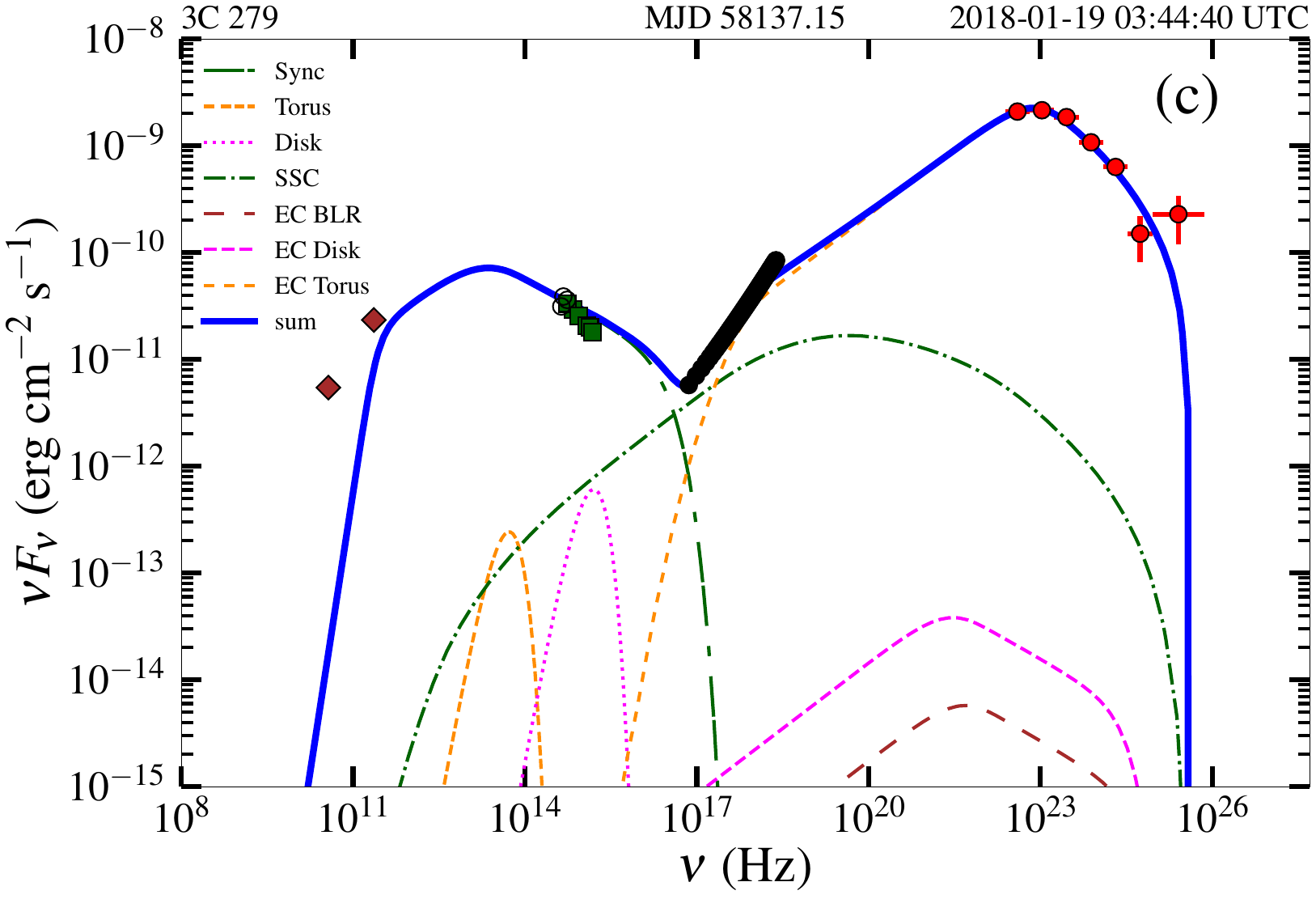}
\includegraphics[width=0.41\textwidth,angle=0]{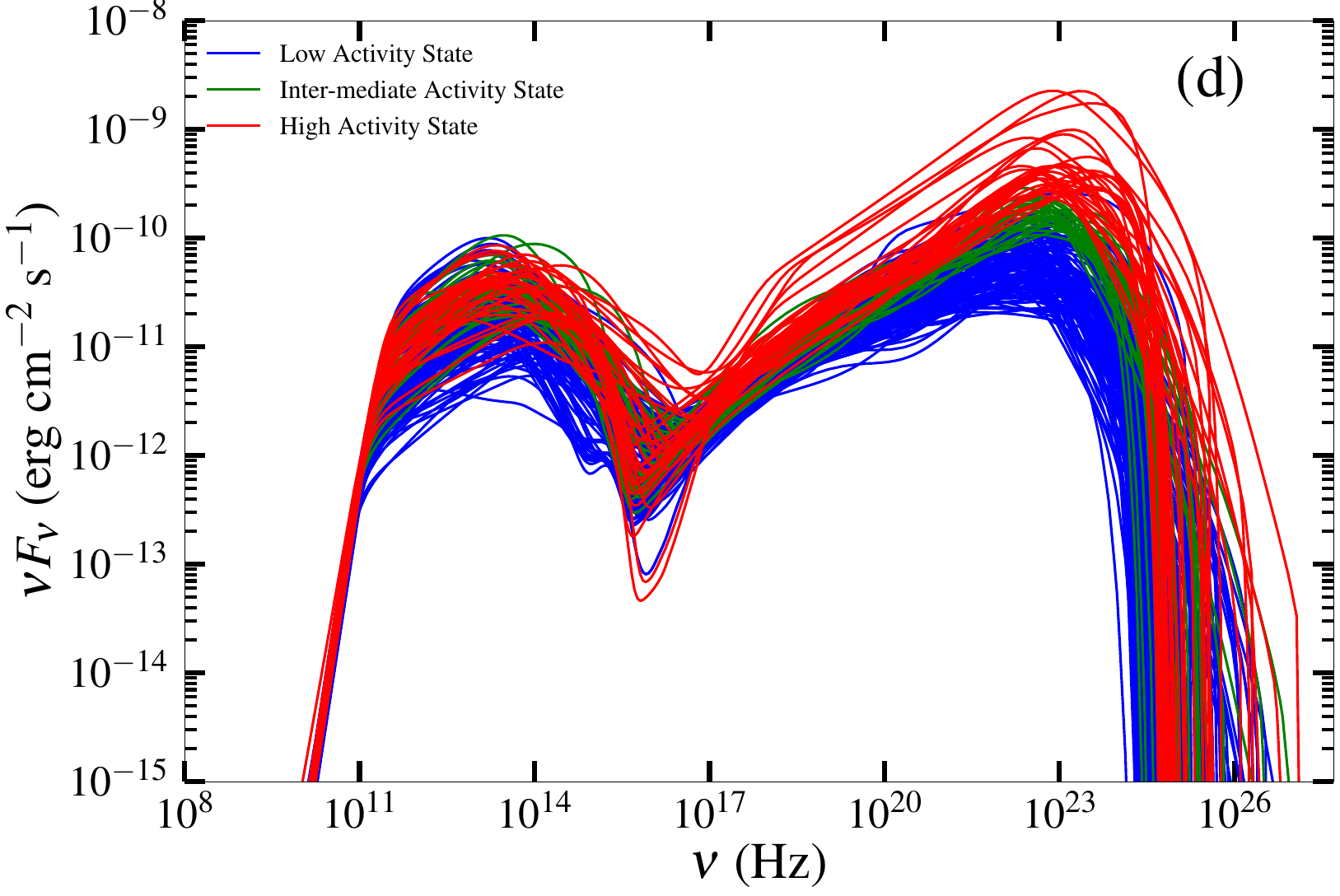}
	\caption{(a) SED fit for quiescent state MJD 56131.65603 -- 56138.65603. (b) SED fit for intermediate state MJD 55662.65603 -- 55669.65603. (c) SED fit for high state MJD 58133.65603 -- 58140.65603. (d) summary of all components from the modeling of all SEDs. Note: This is accompanied by an 84 second video showing the time evolution of SEDs from the activity states (an animation of this figure is available).}
    \label{fig:total_SEDs_all_states}
\end{figure*}

\begin{figure*}
	\centering
	\includegraphics[scale=0.38]{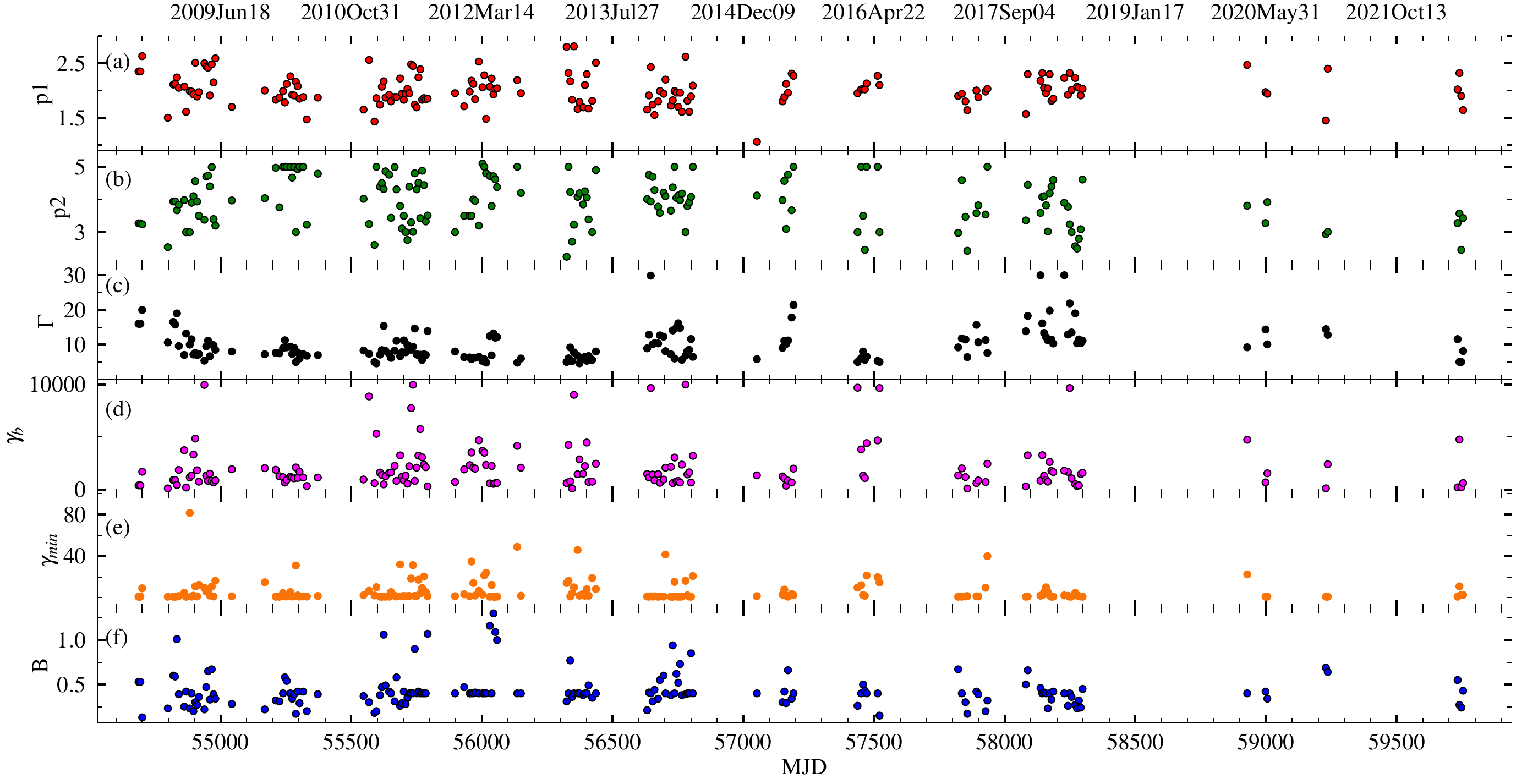}
	\caption{The evolution of model free parameters estimated by modeling the SEDs. Panels (a) and (b): the power-law indexes of electrons before and after the break, Panel (c): the bulk Lorentz factor variation in 2008–2022. Panels (d) and (e): the break and minimum energy of emitting electrons in different epochs. Panel (f): the change of magnetic field in the emitting region.}
    \label{fig4:evolution_of_spectral_free_parameters_SED}
\end{figure*}

\begin{figure}
	\centering
	\includegraphics[scale=0.35]{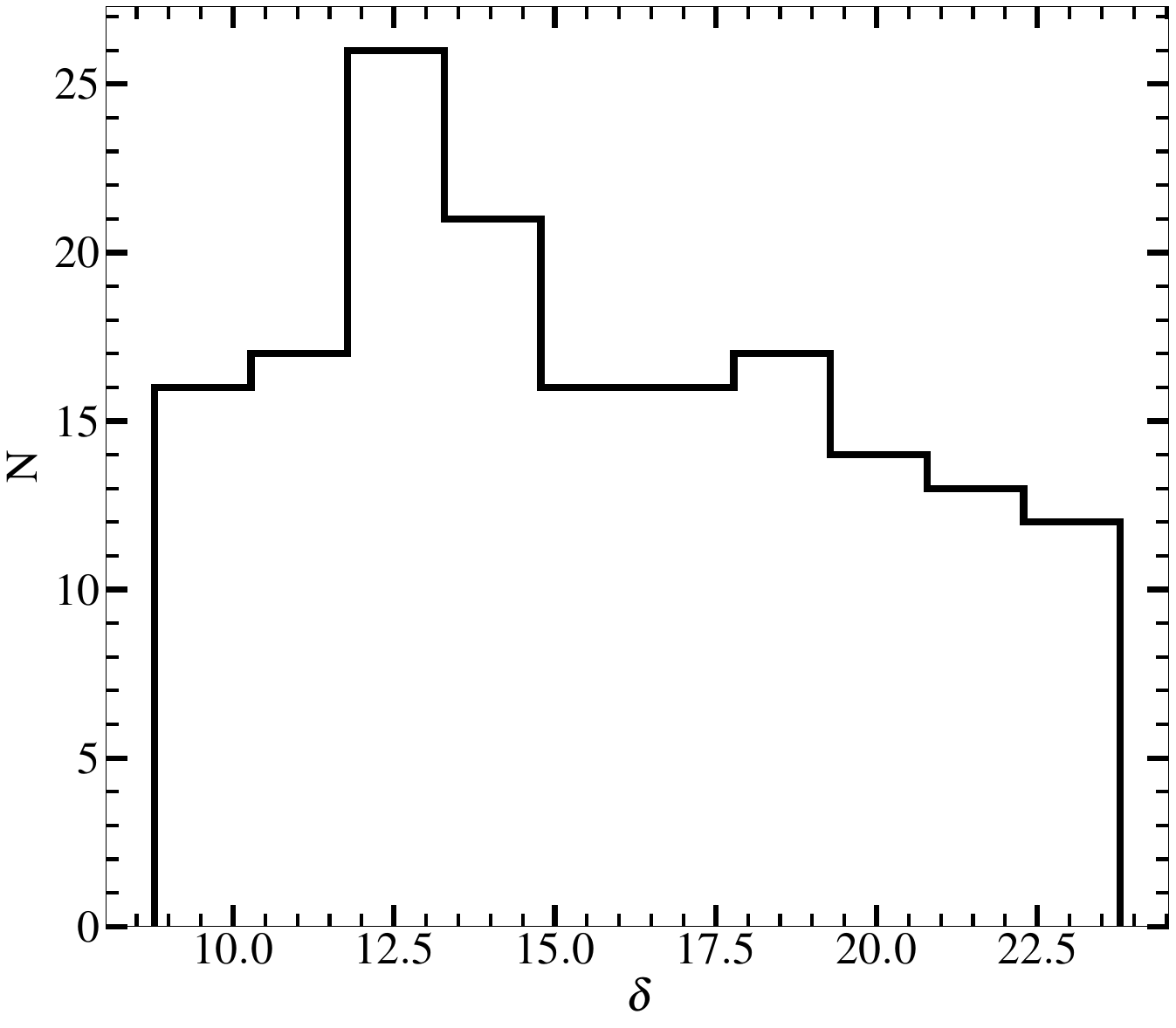}\\
       \includegraphics[scale=0.35]{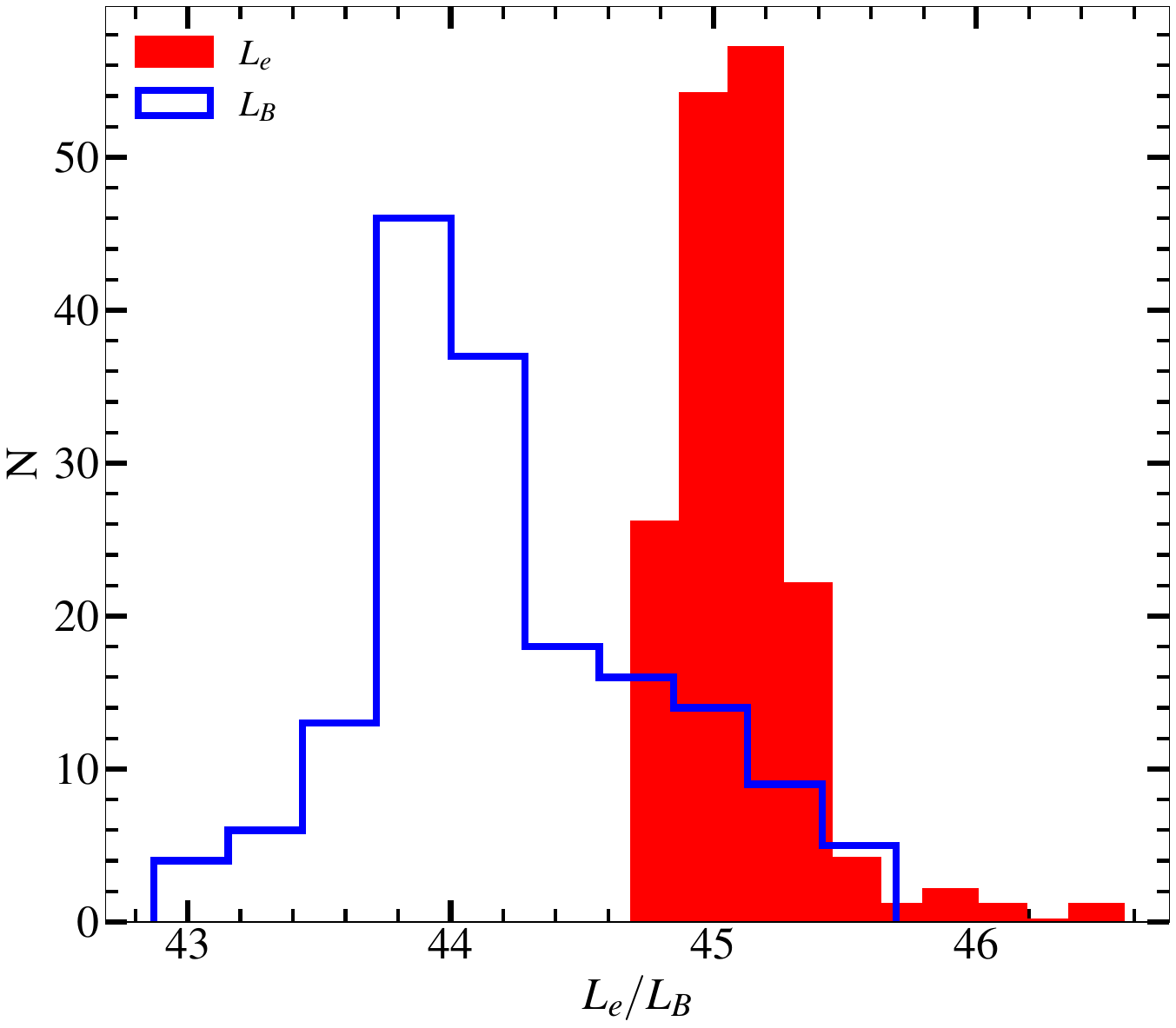}
	\caption{The distribution of Doppler boosting factor. The distribution of the jet luminosity in the form of electrons (filled red) and magnetic field (non filled blue). }
   \label{fig5:delta_Le_LB_distribution}
\end{figure}

The multiwavelength SEDs of 3C 279 in different states during $\sim14$ years of observations
were found to be well modeled with physically realistic parameters within a single-zone leptonic model taking into account the inverse Compton scattering of synchrotron, disk, and BLR photons. 
As an example in Figure~\ref{fig:total_SEDs_all_states} SED modeling for the quiescent state (panel a), intermediate state (panel b), and high state (panel c) is shown. In the same Figure~\ref{fig:total_SEDs_all_states} the derived models for the 168 epochs are presented (panel d). The accompanying video presents a
complete view of the evolution of broadband SEDs over time.
The evolution of the different free parameters of the SEDs is shown in the Figure~\ref{fig4:evolution_of_spectral_free_parameters_SED} and the final parameters are provided in Table~\ref{tab:sed_fit_results}.
From the SED modeling, it was found that the mean values of $p_{\rm 1,mean}=2.01$ and $p_{\rm 2,mean}=3.93$ for all states. For low activity state, the mean value of $p_{\rm 1,mean}=2.04$, $p_{\rm 2,mean}=4.00$; for intermediate activity state, the mean value of $p_{\rm 1,mean}=1.88$, $p_{\rm 2,mean}=3.83$; for flaring activity state $p_{\rm 1,mean}=2.01$, $p_{\rm 2,mean}=3.76$. 
The minimum electron energy ranges from $1.0-81.5$, which is presented in panel (e) of Figure~\ref{fig4:evolution_of_spectral_free_parameters_SED}. The SSC modeling of the X-ray data allows to estimate this minimum electron energy along with $p_{\rm 1}$. The magnetic field was found to vary between $0.13-1.30$ Gauss (panel (f) of Figure~\ref{fig4:evolution_of_spectral_free_parameters_SED}).\\ 

The break energy varies in the range of $100-10000$ for all activity states. For low activity states the break energy varies in the range of $100-10000$; for intermediate activity state the break energy varies in the range of $119-3747$; for flaring activity states the break energy varies in the range of $109-9662$ and is characterized by the interdependence timescales of particle acceleration and cooling. The electron cooling time is defined as 
\begin{equation}
t_{\rm cool}=\frac{3\: m_{e}c \:(1+z)}{4 \sigma_{\rm T}\:u_{\rm tot}^{\prime}\:\gamma_{\rm e}^{\prime}}
\end{equation} 
where $u_{\rm tot}^{\prime}=u_{\rm B}+u_{\rm SSC}+u_{\rm disk}+u_{\rm BLR}$. The densities of photons from the disk and BLR region are constant, however $u_{\rm B}$ and $u_{\rm SSC}$, which depend on the synchrotron and SSC components, vary in different epochs (panel (d) of Figure~\ref{fig:total_SEDs_all_states}). 
Therefore, the small changes in the break energy (see Figure~\ref{fig4:evolution_of_spectral_free_parameters_SED}) is explained with the variation in 
$u_{\rm B}$ and $u_{\rm SSC}$. During most epochs, $u_{\rm SSC}/u_{\rm B}$ was found to be $\geq1$, indicating that the cooling due to SSC is important and thus the nonlinear effects contribute to the particle distribution, i.e., $\gamma_{b}$ and $p_{2}$. In our study, we found that the difference between $p1$ and $p2$ is greater than that expected from the standard cooling break ($\Delta p=1$), which can occur due to the above-mentioned nonlinear effects (for e.g., \citet{sahakyan2021MNRAS}). 
In addition, such a difference in $\Delta p$ could arise due to inhomogeneities or the nature of the injection of particle into the emitting region \citep{reynolds2009ApJ}. \\

The distribution of Doppler boosting factor is presented in the upper panel of Figure~\ref{fig5:delta_Le_LB_distribution}, that has a peak at $\delta \simeq 13$. We noticed that the Doppler boosting factor increases during the intermediate and becomes maximum for the high activity state when compared to low activity states. The highest Doppler boosting factor $\delta =23.79$ was found for the epoch MJD 58245.65603 -- MJD 58252.65603, which is inconsistent with the Doppler beaming value reported by \citet{prince2020ApJ} from SED modeling during the large flare exhibited by the source during April 2018. Also, it is found that for most of the high activity and for few intermediate activity states the Doppler beaming factor is larger compared to the low activity states. This implies a strong Compton dominance during intermediate and flaring activity phases in comparison to the low activity states. This is reflected in the high energy component of the SEDs (Figure~\ref{fig:total_SEDs_all_states}), where the external photon density in the comoving frame depends on the Doppler boosting factor. Such increment in the Doppler boosting factor could due to the re-acceleration of the emitting region during its propagation or emergence of new emitting blob near to the central black hole or the change in the direction of the velocity vector of the emitting plasma \citep{larionov2020}. 
In our study, we noticed during the two of very bright epochs MJD 57188.65603 -- MJD 57195.65603 and MJD 58224.65603 -- MJD 58231.65603, where the emitting region is near to central source when compared to other activity states.
It should be noted that the distance of the emission blob at $15\times10^{17}$ cm could explain the observed SEDs for all the states except for the two very bright states mentioned above. However, for these very bright states, we found that the 
emission blob located at a distance of $3\times10^{17}$ cm could well describe the observed SEDs, which is near the outer boundary of the BLR region. 
An alternative scenario for the increase in Doppler boosting factor could be geometrical effects, such as twisted inhomogeneous jet \citep{raiteri2017Natur} or the jets in a jet \citep{giannios2009MNRAS}. Moreover, the recent work reported by \citet{fuentes2023NatAs} on 3C 279, shows the presence of a twisted filamentary
jet structure and they suggest the enhancement in the Doppler boosting could occur when the emitting region travel in these sites with better alignment with the line of sight, causing the variation in observed flux.   \\

The modeling also provides information on the power of the
jet and these different parameters are provided in Table~\ref{tab:sed_fit_energetics}.
The distribution of the jet luminosities in the form of magnetic
field and electron kinetic energy computed as $L_{B}=\pi c R_b^2 \Gamma^2 U_{B}$ and $L_{e}=\pi c R_b^2 \Gamma^2 U_{e}$ is given in Figure~\ref{fig5:delta_Le_LB_distribution} (lower panel). The mean of $L_{e}$ and $L_{B}$ is at $1.65\times10^{45} {\rm erg\:s^{-1}}$ and $4.50\times10^{44} {\rm erg\:s^{-1}}$, respectively.
$L_{e}$ is in the range of $(0.50-35.50)\times10^{45}{\rm erg\:s^{-1}}$ while $L_{B}$ in $(0.07-49.60)\times10^{44}{\rm erg\:s^{-1}}$. 
The peak of $L_{e}$ is around $1.27\times10^{45}{\rm erg\:s^{-1}}$ and $L_{B}$ is at $7.87\times10^{43}{\rm erg\:s^{-1}}$. For majority of SEDs, we found that the jet is particle dominated 
with $L_{e}/L_{B}\geq1$ and only in a few cases $L_{e}/L_{B}<1$, implying slightly magnetically dominated.
The total jet luminosity estimated as $L_{\rm tot} = L_{\rm e} + L_{\rm B} + L_{\rm p} + L_{\rm rad}$ \citep{ghisellini2001}, where $L_{p}$ and $L_{rad}$ represents the power carried by the cold protons and the radiation produced. We noticed that, except for four high activity states, the total jet luminosity was less than the total Eddington luminosity of $9.93\times10^{46}$ erg sec$^{-1}$ for the black hole mass of $7.9\times10^8\:M_\odot$ in 3C 279. During these four high activity states (MJD 56642.65603 -- 56649.65603, MJD 57181.65603 -- 57188.65603, MJD 58133.65603 -- 58140.65603 and MJD 58224.65603 -- 58231.65603) the total jet power exceeds the disc luminosity $L_{d}$ nearly by an order of magnitude, which is in agreement with the results reported by \citet{ghisellini2014Natur}. 


\section{Summary} \label{sec:summary}
In this work, we performed a systematic construction and modeling of the broad-band
SEDs of the blazar 3C 279 during 2008-2022. We noticed that the source exhibited various activity states like quiescent, intermediate, and flaring states in the $\gamma$-ray band. The multiwavelength SEDs of 3C 279 (in 168 epochs) are constrained with quasi-simultaneous observations in the $\gamma$-ray, X-ray, UV, optical and near-infrared bands.\\

The $\gamma$-ray spectral data points have been analyzed using three distinct spectral models: PL, LP, and BPL. The $TS_{curve}$ derived from all the fittings indicates that during quiescent and intermediate activity states, the $\gamma$-ray spectra is dominated by PL model. However, for high activity state both PL and BPL model were found to be characterizing the $\gamma$-ray photon spectrum. Furthermore, the photon index and flux correlation was investigated in the $\gamma$-ray and X-ray band, which suggests a hint of very moderate harder when
brighter trend. Furthermore, we have calculated the fractional variability across various wavebands, revealing an increase in fractional variability as the energy levels rise.\\

The broadband emission of 3C 279 during different activity states was found to be well described with an assumption of one-zone leptonic scenario taking into account the inverse Compton scattering of synchrotron, disc, BLR and dusty torus reflected photons. It was noticed that the emission region is outside the BLR at a distance of $\sim15\times 10^{17}$ cm from the central black hole. However, to explain the broadband SED during two very bright $\gamma$-ray states, by systematically checking the location of emission region, we found it closer to the outer boundary of the BLR region at a distance of $\sim3\times 10^{17}$ cm from the central black hole. It was noticed that the jet is particle dominated during most of the activity states with only few being magnetically dominant. We infer that during the high activity states the Doppler boosting factor was found to be large compared to the other activity states, which is likely to be caused by the geometrical effects of the jet such as the presence of filament like structure in the jet of 3C 279 as reported by \citet{fuentes2023NatAs}.\\ 

Overall, our study suggests that the multiwavelength emissions from blazar 3C 279 can also be well explained with 
a single-zone model including the synchrotron self-Compton and external Compton scenario. Only for a few of FSRQs and BL Lac objects have detailed long-term SEDs (with a large number of epochs) been reported in the literature. Hence, the validation of different type models in the different class of blazars has a great scope of study. 
Such studies provide a unique opportunity to examine and unveil the dynamical evolution of jet radiation over time due to the collection of numerous high-quality data from observations in various bands.

\section{Acknowledgments}
This work was supported by the CAS `Light of West China' Program (grant No. 2021-XBQNXZ-005), 
the National SKA Program of China (grant No. 2022SKA0120102), and the Tianshan Talent Training Program (grant No. 2023TSYCCX0099). `The CAS `Light of West China' Program (grant No. 2021-XBQNXZ-005)'. KMA acknowledges funding from an ARIES Regular Post-Doctoral Fellowship (AO/RA/2022/1960). JHF's work is partially supported by the National Science Foundation
of China (NSFC 12433004, U2031201, NSFC 11733001).
C.M.R. and M.V. acknowledge financial support from the INAF Fundamental Research Funding Call 2023.
N.S acknowledges the support by the Higher Education and Science Committee of the Republic of Armenia, in the context of the research project No 23LCG-1C004.
We also thank the anonymous reviewer for useful comments which helped us to improve the manuscript.\\

This work has made use of public {\it Fermi}-LAT data obtained from the {\it Fermi} Science Support Center
(FSSC), provided by NASA Goddard Space Flight Center. This work is partly based on data taken and assembled by the WEBT collaboration and stored in the WEBT archive at the Osservatorio Astrofisico di Torino – INAF (\url{https://www.oato.inaf.it/blazars/webt/}).
This work has made use of public data from CSS survey.
The CSS survey is funded by the National Aeronautics and Space
Administration under Grant No. NNG05GF22G issued through the Science
Mission Directorate Near-Earth Objects Observations Program. The CRTS
survey is supported by the U.S.~National Science Foundation under
grants AST-0909182 and AST-1313422.
This paper has made use of up-to-date SMARTS optical/near-infrared light curves that are available at \url{www.astro.yale.edu/smarts/glast/home.php}.
Data from the Steward Observatory spectropolarimetric monitoring project 
were used. This program is supported by Fermi Guest Investigator grants 
NNX08AW56G, NNX09AU10G, NNX12AO93G, and NNX15AU81G.\\

The research at UMRAO was funded in part by a series of grants from the NSF (most recently AST-0607523) and by a series of Fermi G.I. awards from NASA (NNX09AU16G, NNX10AP16G, NNX11AO13G, and NNX13AP18G). Funds for the operation of UMRAO were provided by the University of Michigan. 
The authors thank Margo F. Aller and Hugh D. Aller for providing the UMRAO data.
This study makes use of VLBA data from the VLBA-BU Blazar Monitoring Program (BEAM-ME and VLBA-BU-BLAZAR; \url{http://www.bu.edu/blazars/BEAM-ME.html}), funded by NASA through the Fermi Guest Investigator Program. The VLBA is an instrument of the National Radio Astronomy Observatory. The National Radio Astronomy Observatory is a facility of the National Science Foundation operated by Associated Universities, Inc. The authors thank Svetlana G. Jorstad for providing VLBA data.
The XAO-NSRT is operated by the Urumqi Nanshan Astronomy and Deep Space Exploration Observation and Research Station of Xinjiang (XJYWZ2303). The Submillimeter Array (SMA) is a joint project between the Smithsonian Astrophysical Observatory and the Academia Sinica Institute of Astronomy and Astrophysics and is funded by the Smithsonian Institution and the Academia Sinica. We recognize that Maunakea is a culturally important site for the indigenous Hawaiian people; we are privileged to study the cosmos from its summit. F-GAMMA utilized several facilities in cm, mm, sub-mm, infrared and optical bands, achieving an unprecedented coverage for the study of the spectral evolution of powerful relativistic jets in AGNs. The pivoting facilities were the 100 m radio telescope in Effelsberg (Germany), the 30 m IRAM radio telescope at Pico Veleta (Spain) and the 12-m APEX telescope (Chile). The frequency coverage included 12 frequencies between 2.64 and 345 GHz.
This research has made use of data from the MOJAVE database that is maintained by the MOJAVE team~\citep{lister2018ApJS}.\\


\vspace{5mm}
\facilities{{\sl Fermi}-LAT, {\sl Swift}-XRT/UVOT, SMARTS,  WEBT, KAIT, CRTS, XAO-NSRT, MOJAVE, VLBA-BU Blazar Monitoring Program, Aalto University Mets\"ahovi radio telescope, F-GAMMA, UMRAO, SMA.}
\software{\texttt{Fermitools} version $2.2.0$, \texttt{HEASoft} package (v$6.31$), \texttt{Fermipy} version $1.20$, \texttt{XSpec} version: $12.13.0$ \citep{arnaud1996},  \textit{Jetset} version 1.2.2.}

\appendix
\clearpage
\section{Additional Tables}

\begin{table*}[h]
\caption{Results of $\gamma$-ray analysis with different spectral models PL, LP, and BPL, by using the maximum likelihood analysis method.}
\hspace*{-1.7cm}
\centering
\resizebox{20cm}{!}{
\begin{tabular}{cccccccccc}\hline\hline
Period      &Best fit model  &Flux (10$^{-7}$ photon cm$^{-2}$ s$^{-1}$)   &$\alpha$   &$\beta$   &$\Gamma_{1}$ &$\Gamma_{2}$  &$E_{break}$ (MeV) &$TS$  &$TS_{curve}$ \\
\hline
54682.65603$-$54689.65603    &PL   &$2.95\pm0.56$   &$2.37\pm0.16$    &--  &--  &--   &--   &94.36    &--  \\
54689.65603$-$54696.65603    &PL   &$2.55\pm0.49$   &$2.29\pm0.16$    &--  &--  &--   &--   &110.79    &--  \\
54696.65603$-$54703.65603      &PL   &$17.79\pm0.92$   &$2.17\pm0.04$    &--  &--  &--   &--   &1777.22    &--  \\
54794.65603$-$54801.65603       &PL   &$17.79\pm0.92$   &$2.17\pm0.04$    &--  &--  &--   &--   &1777.03    &--  \\
54815.65603$-$54822.65603       &PL   &$12.59\pm0.69$   &$2.23\pm0.05$    &--  &--  &--   &--   &1547.08    &--  \\
54822.65603$-$54829.65603        &PL   &$8.38\pm3.70$   &$2.18\pm0.11$    &--  &--  &--   &--   &751.23    &--  \\
54829.65603$-$54836.65603       &PL   &$5.97\pm0.77$   &$2.25\pm0.08$    &--  &--  &--   &--   &477.91    &--  \\
\hline
\end{tabular}}
\label{tab:gamma_results}
\end{table*}
\clearpage
\addtocounter{table}{-1}
\begin{table*}[h]
\caption{Continued.}
\hspace*{-1.7cm} 
\centering
\resizebox{20cm}{!}{
\begin{tabular}{cccccccccc}\hline\hline
Period      &Best fit model  &Flux   &$\alpha$   &$\beta$   &$\Gamma_{1}$ &$\Gamma_{2}$  &$E_{break}$ &$TS$  &$TS_{curve}$ \\
\hline
55648.65603$-$55655.65603    &PL   &$3.74\pm0.69$   &$2.41\pm0.16$    &--  &--  &--   &--   &92.42    &--  \\
55662.65603$-$55669.65603    &PL   &$10.10\pm1.52$   &$2.34\pm0.13$    &--  &--  &--   &--   &138.54    &--  \\
55669.65603$-$55676.65603    &PL   &$7.27\pm1.13$   &$2.54\pm0.12$    &--  &--  &--   &--   &331.70    &--  \\
55683.65603$-$55690.65603    &PL   &$5.52\pm0.66$   &$2.36\pm0.10$    &--  &--  &--   &--   &266.75    &--  \\
55690.65603$-$55697.65603    &PL   &$8.18\pm0.87$   &$2.37\pm0.09$    &--  &--  &--   &--   &322.87    &--  \\
55697.65603$-$55704.65603    &PL   &$1.14\pm0.97$   &$2.28\pm0.07$    &--  &--  &--   &--   &589.92    &--  \\
55704.65603$-$55711.65603    &PL   &$8.94\pm0.86$   &$2.10\pm0.07$    &--  &--  &--   &--   &507.99    &--  \\
\hline
\end{tabular}}
\end{table*}
\clearpage
\addtocounter{table}{-1}
\begin{table*}[h]
\caption{Continued.}
\hspace*{-1.7cm} 
\centering
\resizebox{20cm}{!}{
\begin{tabular}{cccccccccc}\hline\hline
Period      &Best fit model  &Flux   &$\alpha$   &$\beta$   &$\Gamma_{1}$ &$\Gamma_{2}$  &$E_{break}$ &$TS$  &$TS_{curve}$ \\
\hline
56642.65603$-$56649.65603    &LP   &$14.62\pm0.89$   &$2.06\pm0.06$    &$0.08\pm0.02$  &--  &--   &--   &1971.86    &11.58  \\
56649.65603$-$56656.65603    &PL   &$19.49\pm0.18$   &$2.32\pm0.04$    &--  &--  &--   &--   &913.65    &--  \\
56656.65603$-$56663.65603    &PL   &$13.31\pm0.99$   &$2.35\pm0.07$    &--  &--  &--   &--   &753.14    &--  \\
56670.65603$-$56677.65603    &PL   &$13.98\pm0.84$   &$2.11\pm0.04$    &--  &--  &--   &--   &1419.25    &--  \\
56677.65603$-$56684.65603    &PL   &$8.08\pm1.28$   &$2.15\pm0.12$    &--  &--  &--   &--   &181.49    &--  \\
56691.65603$-$56698.65603    &PL   &$9.50\pm0.99$   &$2.23\pm0.08$    &--  &--  &--   &--   &377.00    &--  \\
56698.65603$-$56705.65603    &PL   &$5.83\pm0.80$   &$2.23\pm0.10$    &--  &--  &--   &--   &245.01    &--  \\
\hline
\end{tabular}}
\end{table*}

\clearpage
\addtocounter{table}{-1}
\begin{table*}[h]
\caption{Continued.}
\hspace*{-1.7cm} 
\centering
\resizebox{20cm}{!}{
\begin{tabular}{cccccccccc}\hline\hline
Period      &Best fit model  &Flux   &$\alpha$   &$\beta$   &$\Gamma_{1}$ &$\Gamma_{2}$  &$E_{break}$ &$TS$  &$TS_{curve}$ \\
\hline
58245.65603$-$58252.65603      &BPL   &$17.43\pm1.37$   &--    &--  &$1.85\pm0.07$  &$2.35\pm0.13$   &$2000.00\pm0.01$   &$2452.83$    &$10.08$ \\    
58252.65603$-$58259.65603    &PL   &$23.57\pm1.78$   &$1.99\pm0.05$    &--  &--  &--   &--   &$1089.89$    &--  \\  
58266.65603$-$58273.65603    &BPL   &$87.31\pm2.18$   &--    &--  &$1.79\pm0.04$  &$2.54\pm0.09$   &$1103.00\pm159.50$   &$10499.23$    &$93.94$  \\    
58273.65603$-$58280.65603    &PL   &$20.96\pm0.93$   &$2.22\pm0.04$    &--  &--  &--   &--   &$2670.06$    &--  \\    
58280.65603$-$58287.65603    &PL   &$12.09\pm0.56$   &$2.23\pm0.04$    &--  &--  &--   &--   &$1891.05$    &--  \\  
58287.65603$-$58294.65603    &PL   &$29.02\pm1.26$   &$2.16\pm0.04$    &--  &--  &--   &--   &$2608.08$    &--  \\    
58294.65603$-$58301.65603    &PL   &$12.24\pm1.81$   &$2.29\pm0.12$    &--  &--  &--   &--   &$177.69$    &--  \\
\hline
\end{tabular}}
\end{table*}

\clearpage
\begin{table*}[h]
\caption{The variable model parameters estimated by fitting the SEDs for different epochs.}
\hspace*{-1.7cm} 
\centering
\small
\begin{tabular}{ccccccccccc}\hline\hline
Period      &$N_{e}$ (cm$^{-3}$)     &$p_{1}$   &$p_{2}$   &$\gamma_{min}$   &$\gamma_{b}$ &$\Gamma$   &$B$ (G) &$\gamma_{max}$  &$d$ (10$^{17}$ cm)\\
\hline
54682.65603$-$54689.65603    &4982  &2.35   &3.27    &1.0  &398   &16  &0.53  &6259    &15 \\
54689.65603$-$54696.65603    &4982  &2.35   &3.27    &1.0  &398   &16  &0.53  &6259    &15 \\ 
54696.65603$-$54703.65603    &6210  &2.63   &3.24    &9.06  &1700   &19.99  &0.13   &11722  &15 \\
54794.65603$-$54801.65603    &3117  &1.50   &2.54    &1.0  &109   &10.64  &0.23  &8923    &15 \\
54815.65603$-$54822.65603    &2808  &2.11   &3.94    &1.0  &912   &16.54  &0.60  &29562    &15 \\
54822.65603$-$54829.65603    &3205  &2.12   &3.94    &1.0  &910   &15.78  &0.59  &29994    &15 \\
54829.65603$-$54836.65603    &2739  &2.24   &3.67    &1.2  &439   &19  &1.01  &4521    &15 \\
\hline
\end{tabular}
\label{tab:sed_fit_results}
\end{table*}

\clearpage
\addtocounter{table}{-1}
\begin{table*}[h]
\caption{continued.}
\hspace*{-1.7cm} 
\centering
\small
\begin{tabular}{ccccccccccc}\hline\hline
Period      &$N_{e}$ (cm$^{-3}$)     &$p_{1}$   &$p_{2}$   &$\gamma_{min}$   &$\gamma_{b}$ &$\Gamma$   &$B$ (G) &$\gamma_{max}$  &$d$ (10$^{17}$ cm)\\
\hline
55697.65603$-$55704.65603     &1977  &1.83   &3.50    &1.59  &894   &11.21  &0.42  &4584    &15 \\
55704.65603$-$55711.65603     &6352  &1.94   &3.0    &1.54 &1318   &7.79  &0.28  &10550    &15 \\
55711.65603$-$55718.65603     &6584  &2.03   &2.76    &1.48  &566   &9.86  &0.35  &7075   &15 \\
55718.65603$-$55725.65603     &3412  &1.95   &4.39    &1.65  &2218   &9.0  &0.40  &345644    &15 \\
55725.65603$-$55732.65603     &1245  &2.48   &3.30    &18.49  &7748   &8.50  &0.40  &6221    &15 \\
55732.65603$-$55739.65603     &490  &2.45   &3.01   &31.28  &9979   &9.40  &0.40  &4938   &15 \\
55739.65603$-$55746.65603     &497  &1.74   &4.80    &1.77  &833   &14.65  &0.90  &59445    &15 \\
\hline
\end{tabular}
\end{table*}

\clearpage
\addtocounter{table}{-1}
\begin{table*}[h]
\caption{continued.}
\hspace*{-1.7cm} 
\centering
\small
\begin{tabular}{ccccccccccc}\hline\hline
Period      &$N_{e}$ (cm$^{-3}$)    &$p_{1}$   &$p_{2}$   &$\gamma_{min}$   &$\gamma_{b}$ &$\Gamma$   &$B$ (G) &$\gamma_{max}$  &$d$ (10$^{17}$ cm)\\
\hline
56691.65603$-$56698.65603     &3010  &1.94   &4.21    &1.10  &957   &12.30  &0.60  &26611    &15 \\
56698.65603$-$56705.65603     &318  &2.20   &4.10    &41.56  &2081   &8.10  &0.4  &66562    &15 \\
56719.65603$-$56726.65603     &2310  &1.72   &3.66    &1.00  &2155   &7.20  &0.38  &34028    &15 \\
56726.65603$-$56733.65603     &1268  &1.82   &4.37    &1.01  &619   &14.10  &0.94  &19474    &15 \\
56733.65603$-$56740.65603     &1020  &1.99   &5.00    &15.23  &3051   &6.0  &0.4  &56009    &15 \\
56740.65603$-$56747.65603     &3100  &1.96   &4.05    &1.15  &778   &14.90  &0.62  &458157    &15 \\
56747.65603$-$56754.65603     &1153  &1.70   &4.12    &1.02  &822   &16.15  &0.52  &17011    &15 \\
\hline
\end{tabular}
\end{table*}

\clearpage
\addtocounter{table}{-1}
\begin{table*}[h]
\caption{continued.}
\hspace*{-1.7cm} 
\centering
\small
\begin{tabular}{ccccccccccc}\hline\hline
Period      &$N_{e}$ (cm$^{-3}$)    &$p_{1}$   &$p_{2}$   &$\gamma_{min}$   &$\gamma_{b}$ &$\Gamma$   &$B$ (G) &$\gamma_{max}$  &$d$ (10$^{17}$ cm)\\
\hline
58287.65603$-$58294.65603     &5454  &1.91   &3.09    &1.23  &1469   &10.41  &0.24  &9175    &15 \\
58294.65603$-$58301.65603     &5817  &2.03   &4.61    &1.01  &1594  &11.14  &0.45  &12721    &15 \\
58924.65603$-$58931.65603     &719  &2.47   &3.81    &22.46  &4741   &9.22 &0.4  &8237    &15 \\
58994.65603$-$59001.65603     &2913  &1.97   &3.28    &1.00  &686   &14.36  &0.42  &7635    &15 \\
59001.65603$-$59008.65603     &5493  &1.94   &3.92    &1.13  &1550  &10.08  &0.34  &9983    &15 \\
59225.65603$-$59232.65603     &1000  &1.45   &2.94    &1.00  &119   &14.46 &0.69  &8802    &15 \\
59232.65603$-$59239.65603     &10000  &2.40   &3.01    &1.0  &2406   &12.80  &0.64  &5149    &15 \\
59729.65603$-$59736.65603     &8320  &2.02   &3.28    &1.02  &217   &11.58  &0.55  &6122    &15 \\
59736.65603$-$59743.65603     &5314  &2.32   &3.57    &10.98  &4765   &5.0  &0.27  &13674    &15 \\
59743.65603$-$59750.65603     &9993  &1.90   &2.46    &2.75  &227   &5.0  &0.24  &11409    &15 \\
59750.65603$-$59757.65603     &1007  &1.64   &3.43    &2.75  &621   &8.16  &0.43  &8163    &15 \\
                            &  &   &    &  &   &  &  &    & \\
                            
\hline
\end{tabular}
\end{table*}

\begin{table*}[h]
\caption{Energy densities, jet luminosities parameters derived from fitted SED parameters.}
\hspace*{-1.7cm} 
\centering
\resizebox{20cm}{!}{
\begin{tabular}{ccccccccccc}\hline\hline
Period      &$U_{e}$ (erg cm$^{-3}$)     &$U_{B}$ (erg cm$^{-3}$)   &$\eta$ ($U_{e}/U_{B}$) &$U_{p}$ (erg cm$^{-3}$) &$L_{rad}$ (erg s$^{-1}$)  &$L_{kin}$ (erg s$^{-1}$)   &$L_{tot}$ (erg s$^{-1}$) &$L_{e}$ (erg s$^{-1}$) &$L_{B}$ (erg s$^{-1}$) &$L_{p}$ (erg s$^{-1}$)\\
\hline
54682.65603$-$54689.65603    &1.43e-02 &1.12e-02 &1.28 &7.49e-01  &1.49e+44 &6.61e+46 &6.72e+46 &1.24e+45 &9.68e+44 &6.49e+46 \\ 
54689.65603$-$54696.65603    &1.43e-02 &1.12e-02 &1.28 &7.49e-01  &1.49e+44 &6.61e+46 &6.72e+46 &1.24e+45 &9.68e+44 &6.49e+46 \\
54696.65603$-$54703.65603    &1.17e-01  &6.72e-04  &174.11    &9.33e-01 &2.06e+45 &3.20e+47 &6.62e+46 &3.55e+46 &2.05e+44 &2.84e+46 \\
54794.65603$-$54801.65603    &4.96e-02 &2.10e-03 &23.59    &4.68e-01 &6.67e+44
 &1.98e+46 &2.05e+46 &1.90e+45 &8.04e+43 &1.79e+46 \\
54815.65603$-$54822.65603    &1.28e-02 &1.43e-02 &0.895   &4.22e-01    &4.34e+44 &4.03e+46 &4.20e+46 &1.19e+45 &1.33e+45 &3.91e+46 \\
54822.65603$-$54829.65603    &1.43e-02 &1.38e-02 &1.03    &4.82e-01 &3.94e+44 &4.18e+46
&4.34e+46 &1.21e+45 &1.17e+45 &4.06e+46 \\
54829.65603$-$54836.65603    &1.10e-02  &4.06e-02   &0.27   &4.12e-01  &3.38e+44 &5.17e+46 &5.70e+46 &1.34e+45 &4.96e+45 &5.03e+46 \\
\hline
\end{tabular}}
\label{tab:sed_fit_energetics}
\end{table*}

\addtocounter{table}{-1}
\begin{table*}[h]
\caption{Continued.}
\hspace*{-1.7cm} 
\centering
\resizebox{20cm}{!}{
\begin{tabular}{ccccccccccc}\hline\hline
Period      &$U_{e}$ (erg cm$^{-3}$)     &$U_{B}$ (erg cm$^{-3}$)   &$\eta$ ($U_{e}/U_{B}$) &$U_{p}$ (erg cm$^{-3}$) &$L_{rad}$ (erg s$^{-1}$)  &$L_{kin}$ (erg s$^{-1}$)   &$L_{tot}$ (erg s$^{-1}$) &$L_{e}$ (erg s$^{-1}$) &$L_{B}$ (erg s$^{-1}$) &$L_{p}$ (erg s$^{-1}$)\\
\hline
55760.65603$-$55767.65603  &8.76e-02 &6.37e-03 &13.75 &1.17 &2.10e+44 &2.12e+46
&2.15e+46 &1.48e+45 &1.08e+44 &1.97e+46 \\
55767.65603$-$55774.65603  &5.62e-02  &6.37e-03   &8.82    &1.21e-01 &4.40e+44 &1.86e+45 &2.37e+45 &5.88e+44 &6.66e+43 &1.27e+45 \\
55774.65603$-$55781.65603  &3.12e-02 &6.37e-03 &4.90  &4.27e-02  &3.37e+44 &1.28e+45
&1.73e+45 &5.43e+44 &1.11e+44 &7.43e+44 \\
55781.65603$-$55788.65603  &4.59e-02 &6.37e-03 &7.20  &1.58e-01 &5.22e+44 &3.36e+45
&3.99e+45 &7.56e+44 &1.05e+44 &2.60e+45 \\
55788.65603$-$55795.65603  &1.10e-02 &4.55e-02 &0.24    &1.52e-01 &2.71e+44 &1.06e+46 &1.39e+46 &7.20e+44 &2.97e+45 &9.90e+45 \\
55893.65603$-$55900.65603  &4.69e-02 &6.37e-03 &7.36  &7.21e-01 &2.56e+44 &1.65e+46
&1.69e+46 &1.01e+45 &1.37e+44 &1.55e+46 \\
55928.65603$-$55935.65603  &5.10e-02 &8.79e-03 &5.80  &1.80e-01 &5.75e+44 &3.17e+45 &3.87e+45 &6.99e+44 &1.20e+44 &2.47e+45 \\
\hline
\end{tabular}}
\end{table*}

\addtocounter{table}{-1}
\begin{table*}[h]
\caption{Continued.}
\hspace*{-1.7cm} 
\centering
\resizebox{20cm}{!}{
\begin{tabular}{ccccccccccc}\hline\hline
Period      &$U_{e}$ (erg cm$^{-3}$)     &$U_{B}$ (erg cm$^{-3}$)   &$\eta$ ($U_{e}/U_{B}$) &$U_{p}$ (erg cm$^{-3}$) &$L_{rad}$ (erg s$^{-1}$)  &$L_{kin}$ (erg s$^{-1}$)   &$L_{tot}$ (erg s$^{-1}$) &$L_{e}$ (erg s$^{-1}$) &$L_{B}$ (erg s$^{-1}$) &$L_{p}$ (erg s$^{-1}$)\\
\hline
57167.65603$-$57174.65603  &2.98e-02  &1.73e-02 &1.72  &6.61e-01 &3.38e+44 &2.92e+46 &3.03e+46 &1.26e+45 &7.33e+44 &2.79e+46 \\
57181.65603$-$57188.65603  &6.46e-02 &4.60e-03 &14.04  &9.03e-01 &1.92e+45 &1.04e+47
&1.06e+47 &6.96e+45 &4.95e+44 &9.72e+46 \\
57188.65603$-$57195.65603  &2.57e-02 &6.37e-03 &4.03    &4.72e-01 &2.41e+45 &7.76e+46 &8.10e+46 &4.01e+45 &9.93e+44
&7.36e+46 \\
57433.65603$-$57440.65603  &1.22e-01 &2.78e-03 &43.88    &3.02e-01 &1.04e+45
&3.53e+45 &4.59e+45 &1.02e+45 &2.30e+43 &2.51e+45 \\
57447.65603$-$57454.65603  &4.62e-02 &6.37e-03 &7.25   &1.20e-01 &3.30e+44
 &2.08e+45 &2.49e+45 &5.79e+44 &7.97e+43 &1.51e+45 \\
57454.65603$-$57461.65603  &3.86e-02 &9.95e-03 &3.88    &4.38e-01 &2.60e+44 &1.02e+46
  &1.07e+46 &8.31e+44 &2.14e+44 &9.42e+45 \\
57461.65603$-$57468.65603  &8.98e-02 &7.35e-03  &12.22  &1.19 &4.28e+44 &1.34e+46
 &1.39e+46 &9.39e+44  &7.70e+43  &1.25e+46 \\
\hline
\end{tabular}}
\end{table*}
\bibliography{references}{}
\bibliographystyle{aasjournal}



\end{document}